\documentstyle{mn}

\def\today{\ifcase\month\or
 January\or February\or March\or April\or May\or June\or
 July\or August\or September\or October\or November\or
 December\fi\space\number\day, \number\year}

\def\todmy{\number\day\space\ifcase\month\or
 January\or February\or March\or April\or May\or June\or
 July\or August\or September\or October\or November\or
 December\fi\space\number\year}

\newcommand{\beqn} {\begin{equation}}
\newcommand{\eeqn} {\end{equation}}
\newcommand{\beqr} {\begin{array}}
\newcommand{\eeqr} {\end{array}}

\newcommand{\AAA}{{A\&A}}

\newcommand{\ApJ}{{ApJ}}

\newcommand{\ApJS}{{ApJS}}
\newcommand{\AJ}{{AJ}}

\newcommand{\MN}{{MNRAS}}

\newcommand{\PASJ}{{PASJ}}

\title{What determines the strength and the slowdown rate of bars ?} 
\author[E. Athanassoula]
       {
       E. Athanassoula \\
Observatoire de Marseille, 
2 Place Le Verrier, 
F-13248 Marseille Cedex 4, France \\
}

\date{Accepted .
      Received ;
      }

\pagerange{\pageref{firstpage}--\pageref{lastpage}}
\pubyear{2002}

\begin{document}

\maketitle

\label{firstpage} 
\begin{abstract}
Isolated barred galaxies evolve by redistributing their angular
momentum, which, emitted by material in the inner disc at resonance 
with the bar, can be absorbed by resonant material in the outer disc, or
in the halo. The amount of angular momentum that can be
emitted/absorbed at a given resonance depends on the distribution
function of the emitting/absorbing material. It thus depends not only
on the amount of material on resonant orbits, but also on the velocity 
dispersion of that material. As it loses angular momentum, the bar
becomes stronger and it also rotates slower. Thus the strength of the
bar and the decrease 
of its pattern speed with time are set by the amount of angular
momentum exchanged within the galaxy, which, in turn, is regulated by
the mass distribution and the velocity dispersion of the 
material in the disc and spheroidal components. Correlations between the
pattern speed of the bar, its strength and the angular momentum
absorbed by the spheroid (halo plus bulge)
argue strongly that it is the amount of angular momentum exchanged
that determines the strength and the slowdown rate of the bar. The
decrease of the bar pattern speed with time should not be used to set
constraints on the halo-to-disc mass ratio, since it depends also on
the velocity dispersion of the halo and disc material. 
\end{abstract}


\section{Introduction}
\indent

Leafing through an atlas of galaxies (e.g. Sandage 1961, Sandage \&
Bedke 1988) or looking at images of barred galaxies on different web sites one
sees that bars come in a variety of shapes and sizes. From the strong
bars, like in NGC 1365, NGC 1300 and NGC 5383, to the small bars confined
to the central parts, like in our own Galaxy, and to the ovals, like 
in NGC 1566, all possible lengths and strengths are covered. Several studies
have been devoted to finding some systematic trends in their
properties. Athanassoula \& Martinet (1980) and Martin (1995) found a correlation
between the length of the bar and the size of the bulge components,
while Elmegreen \& Elmegreen (1985) showed that 
there is a clear dependence of the bar length
on the galaxy type, in the sense that early type disc galaxies have
longer bars than late types. Similarly, Fourier analysis of
the deprojected light distribution of barred galaxies,
shows (Ohta 1996 and references therein) that early types have
considerably higher values for their $m$ 
= 2, 4, 6 and 8 Fourier components. These works, and several others,
point to the fact that bars in early type disc galaxies are on average
stronger than those in later types.  

Bar pattern speeds are much more difficult to measure. Observational determinations are
either indirect (e.g. from the location of rings, or from fits of
gas flow models to velocity field data), or with the help of
the Tremaine \& Weinberg (1984b) method. The value of the pattern
speed at a given time is a function of its initial value, as well as
of its change during the evolution. The latter of course is not possible to
determine observationally, until observations of far away galaxies
become sufficiently detailed to allow it to be estimated.

So what determines the strength of a bar and the evolution of its
pattern speed? In this paper I will use both analytical calculations and
$N$-body simulations to argue that it is the angular
momentum exchange between different parts of a galaxy. 
The role of the angular momentum exchange has already been discussed
in many papers. First and foremost, the ground-braking paper of
Lynden-Bell \& Kalnajs (1972, hereafter LBK), who argue that 
angular momentum exchange is the mechanism that generates
spirals. Mark (1976) discussed wave amplification through processes
that remove angular momentum from galactic discs, while Kormendy
(1979) proposed that the 
angular momentum exchange between the bar and the spheroid could drive
secular evolution. Sellwood (1980) was the first to measure in an
$N$-body simulation the angular momentum exchange between the disc and
the halo component. Tagger et al. (1987) and Sygnet et al. (1988)
showed that
mode coupling allows a galaxy to transfer angular momentum over a
larger radial interval than what a single mode would have allowed. The
link between angular momentum and 
bar slowdown has been made in many papers (Tremaine \& Weinberg 1984a;
Weinberg 1985; Little \& Carlberg 1991a, 1991b; Hernquist \& Weinberg
1992, Athanassoula 1996; Debattista \& Sellwood 1998, 2000; Valenzuela
\& Klypin 2002), out of which some calculated, or at least emphasized,
the role of resonant stars (Tremaine \& Weinberg 1984a,
Weinberg 1985, Little \& Carlberg 1991a, Hernquist \& Weinberg
1992). The link between the angular momentum exchange and the bar
strength has been addressed only recently 
(Athanassoula 2002a, hereafter A02). Although all these papers show a general
qualitative agreement, in the sense that bars grow stronger and slow
down as they evolve, quantitatively they disagree. Not much effort has
been put in understanding the reason for these differences, which were
even some times attributed to inadequacies of either the codes or
the models. 

In this paper, I will argue that it is the angular momentum exchange
within the galaxy that determines the bar growth and its slowdown. I
will seek what influences the angular momentum exchange and therefore
the bar strength and slowdown rate. In section \ref{sec:analytic}, I
will present some, mainly linear, 3D theoretical work, which I will apply to
the disc and to the spheroidal (halo and bulge) components
separately. This will allow me to determine which parts of the galaxy
gain and which parts lose angular momentum and also to get some insight 
about the quantities that can influence the amount exchanged. In section
\ref{sec:orbits}, I will discuss resonances and their link to the disc
orbital structure. Section~\ref{sec:simulations} introduces the
simulations and section \ref{sec:angmom} discusses the angular
momentum exchange. Sections \ref{sec:discdisp}, \ref{sec:density}, and
\ref{sec:halodisp} discuss the effect of the disc-to-halo mass ratio
and of the velocity dispersions. Section
\ref{sec:correlations} presents correlations, based on the results of a
very large number of simulations, that establish the link between the
angular momentum exchange and the bar strength and pattern
speed. Finally, in section \ref{sec:sumdisc}, I summarise and present a
general discussion, including the applicability of the simulation
results to real galaxies and the conclusions that can be reached.
  
\section{Analytical calculations}
\label{sec:analytic}

Using linear theory, it is possible to follow the angular momentum
redistribution. For this I will follow the method already outlined in
several papers (LBK, Kato 1971,
Weinberg \& Tremaine 1984a, Weinberg 1985 etc.). I will thus use
Hamilton-Jacobi variables ($J_i$,$w_i$), where $J_i$ are the canonical
momenta, $w_i$ are the angles and $i$ = 1, 2, 3. The equations of motion
of a given particle are then

\begin{equation}
    \dot J_{i} =  -\frac{\partial H_{0}}{\partial w_i}=0,
    ~~~~~~\dot w_i  =  \frac{\partial H_{0}}{\partial J_i}\equiv
     \Omega_{i},
\label{eq:motion}
\end{equation}

\noindent
where $\Omega_i$ are the frequencies and $H_0$ is the unperturbed
Hamiltonian. The potential $\Psi$ can be written as a sum of an
axisymmetric component $\Psi_0$ and a perturbation, i.e.

\begin{equation}
\Psi = \Psi_0 + \psi e^{i\omega t},
\end{equation}

\noindent
where $\omega$ is the wave frequency. Its real part $\omega_R$ gives
the pattern speed, $\omega_R$ = $\Omega_p / m$, and its imaginary part, $\omega_I$,
gives the growth rate. 
Since the angle variables are periodic in phase space with period
2$\pi$,the potential perturbation  can be expanded in a Fourier series. 

\begin{equation}
     \psi (J_i, w_i) =  \frac{1}{8\pi^3}
     \sum_{l,m,n}\psi_{lmn}(J_i)\ e^{i(lw_1 + mw_2 + nw_3)}.
\end{equation}

\noindent
The coefficients $\psi_{lmn}$ are given by

\begin{eqnarray}
\psi_{lmn}(J_i) = \int_0^{2\pi} \int_0^{2\pi} \int_0^{2\pi} dw_1 dw_2
dw_3~~~~~~~~~~~~~~~~~~~~  \nonumber \\
\times~~\psi (J_i, w_i)~e^{-i(lw_1+mw_2+nw_3)}.
\end{eqnarray}

\noindent
Similar equations can be written for the perturbation of the
distribution function and of the density.

LBK calculated the change of an orbit due to the potential
perturbation. To first order one can calculate the forces along the
unperturbed orbit. In this case, the changes in the actions
$\Delta_1J_i$ are periodic in the angle variables $w_i$. Therefore, for
particles distributed uniformly in $w_i$, the average gain in angular
momentum\footnote{in the following I will often loosely refer to the $z$
  component of the angular momentum as the angular momentum.} (obtained
by integrating over $w_i$) is zero, i.e. particles can 
neither give nor take angular momentum. Thus the change is of second order
and has to be obtained by calculating the forces along the perturbed
orbit. To obtain the total change in angular momentum one has to 
integrate over the unperturbed distribution function $F$. Following
LBK, I get

\begin{eqnarray} 
\dot L_z = \frac {1}{8\pi^2} \omega_I e^{-2\omega_It}
      \int \int \int dJ_1 dJ_2 dJ_3~~~~~~~~~~~~~~~~~~~~   \nonumber \\
\times~~\sum_{l,m,n}\
      \frac{m \left(l\frac{\partial F}{\partial J_1}
+m\frac{\partial F}{\partial J_2}
+n\frac{\partial F}{\partial J_3}\right)}
{|l\Omega_1+m\Omega_2+n\Omega_3+\omega|^2}\ |\psi_{lmn}|^2 ,
\label{eq:DLgeneral1}
\end{eqnarray}

\noindent
where $L_z$ is the $z$ component of the angular momentum, and the 
integration is carried over all the available action space. If
$\omega_I$ = 0, then the integral on the right side is non-zero only
at the resonances, i.e. where

\begin{equation}
l\Omega_1+m\Omega_2+n\Omega_3 = - \omega_R = m \Omega_p 
\label{eq:resonances}
\end{equation}

\noindent
On the other hand, in the
case of a non-zero $\omega_I$, this integral is non-zero even away from
resonances, i.e. angular momentum can be emitted or absorbed even away
from resonances. This contribution, however, will be small if the value
of $\omega_I$ is not large. For $\omega_I \rightarrow 0$ one can write

\begin{eqnarray} 
\dot L_z = - \frac {1}{8\pi} 
      \int \int \int dJ_1 dJ_2 dJ_3~~~~~~~~~~~~~~~~~~~~~~~~~~~~~~~  
      \nonumber  \\
      \times~~\sum_{l,m,n}\
      m \left (l\frac{\partial F}{\partial J_1}
+m\frac{\partial F}{\partial J_2}
+n\frac{\partial F}{\partial J_3}\right)
      \nonumber \\
      \times~~|\psi_{lmn}|^2 
\delta (l\Omega_1+m\Omega_2+n\Omega_3+\omega).
\label{eq:DLgeneral2}
\end{eqnarray}

\noindent
This equation is particularly useful if one wishes to find which
resonances emit angular momentum and which absorb it, and will be
discussed later, separately for the disc and spheroidal
components. Integrating eq. (\ref{eq:DLgeneral1}) over time I can find the
total change of angular momentum 

\begin{eqnarray} 
\Delta L_z = - \frac {1}{16\pi^2} e^{-2\omega_It}
      \int \int \int dJ_1 dJ_2 dJ_3~~~~~~~~~~~~~~~~~ 
      \nonumber  \\
      \times~~\sum_{l,m,n}\
      \times~~\frac{m \left (l\frac{\partial F}{\partial J_1}
+m\frac{\partial F}{\partial J_2}
+n\frac{\partial F}{\partial J_3}\right)}
{|l\Omega_1+m\Omega_2+n\Omega_3+\omega|^2}\ |\psi_{lmn}|^2.
\label{eq:DLgeneral}
\end{eqnarray}

\noindent
Thus linear theory predicts a direct relationship between
the perturbing potential $\psi$ and the change of angular momentum
$\Delta L_z$. In the simple case of an external forcing due to a
companion (e.g. Berentzen, Athanassoula, Heller et al. 2003) or to a rigid
bar (e.g. Weinberg 1985, Hernquist and Weinberg 1992) on a given target disc, 
this equation will imply a correlation between the change in angular
momentum and the 
amplitude of the external forcing. In the more general case, however,
of the self consistent evolution of different disc/halo
configurations, as those considered in the present paper, the relation
between the bar strength and the change of angular momentum, although
straightforward, is not an exact correlation, since different cases
will have different distribution functions and different forms of
perturbation potentials. 

The above equations make no predictions about the pattern speed. This,
in the case of modes, is set by the mode conditions, and, in cases with
external forcing, by the pattern speed of the forcing. Thus, to find 
the relation between the exchange of angular momentum and the pattern
speed, one should apply nonlinear theory. For the bar component,
considered as a solid body, I can write  

\begin{equation}
L_{z,B} = I~\Omega_p,
\label{eq:Lomp}
\end{equation}    
 
\noindent
where $L_{z,B}$ is the bar angular momentum, $\Omega_p$ is the pattern
speed of the bar and $I$ is its moment of inertia. If most of the angular
momentum lost by the bar was taken by the spheroid,
then I can assume that the angular momentum of the outer part of the
disc does not change with the evolution and thus I can, for this case, write  
$L_{D,inner} - I~\Omega_p = L_S$, where $L_{D,inner}$ is the angular momentum
of the inner disc initially and $L_S$ is the angular momentum taken by
the spheroid, which, if the spheroid was initially non-rotating, is
just the total angular momentum of the halo and bulge components at the
time under consideration. Thus in such a case I would expect, at any
given time, a simple linear relation between the bar pattern speed and
the angular momentum taken by the spheroid.

The evolution of the bar angular momentum can be given as

\begin{equation}
\frac {dL_{z,B}} {dt} = \frac {d(I~\Omega_p)} {dt}.
\label{eq:Ldomp}
\end{equation}    
 
\noindent
The change of angular momentum will thus depend not only on the change
of $\Omega_p$, but also on the change of the moment of inertia. 
One thus expects the change of the
angular momentum to be directly proportional to the change of the bar
pattern speed only if the momentum of inertia of the bar does not
change, i.e. if the bar is rigid (as e.g. for Weinberg 1985, or
Hernquist and Weinberg 1992). This is not true in a general case where
the bar evolves self-consistently, since both the bar pattern
speed and its moment of inertia change with time.   

Eq.~(\ref{eq:DLgeneral}) can be somewhat simplified if one considers
the disc and halo components separately. 

\subsection{Disc component}

For the disc component one can
neglect the $z$ dimension and obtain eq. (29) of LBK. In the epicyclic
approximation 

\begin{equation}
    \Omega_1 =  \kappa,~~~~~\Omega_2 = \Omega, \\
\label{eq:epicdefO} 
\end{equation}
\begin{equation}
    J_1 = \frac{1} {2} \kappa a^2, ~~~~~J_2 = L_z,
\label{eq:epicdefJ} 
\end{equation}

\noindent
where $\Omega$ and $\kappa$ are the angular and epicyclic frequencies
respectively and $a$ is the amplitude of the epicyclic oscillation. 
Furthermore, in this approximation 

\begin{equation}
|\frac{\partial F}{\partial J_1}| 
>> |\frac{\partial F}{\partial J_2}|, 
\label{eq:inequality} 
\end{equation}

\noindent
so that for $l$ different from
0 one may retain in eq.~(\ref{eq:DLgeneral2}) only the first term of
the quantity in parenthesis, i.e. $l\frac{\partial F}{\partial J_1}$.
Since $\frac{\partial F}{\partial J_1} < 0$ for any sensible
distribution function, the sign of the product $lm$ will determine whether
angular momentum is gained or lost at the resonance defined by those
$l$ and $m$. For $l$ = 0 the first term vanishes, so that $\dot L_z$
will always be positive. In general, however, the contribution of this
resonance will in general be absolutely
smaller than the corresponding terms for $l$ different than zero,
because of inequality (\ref{eq:inequality}).

By recasting eq.~(\ref{eq:DLgeneral2}), LBK showed that the
perturbation has negative energy and angular momentum within 
corotation (see also Kalnajs 1971). This means that, if energy or
angular momentum is given to 
it, it will be damped, while if is taken from it it will be excited.
  
If I use the standard form of the disc distribution function

\begin{equation}
F_0(J_1, J_2)=\frac{1}{(2\pi)^2<J_1>}~~\Sigma_d (J_2)~~e^{-J_1/<J_1>},
\end{equation}

\noindent
I see that $\frac{\partial F}{\partial J_1}$ is equal
to $ -F / <J_1>$, or, using eq.~(\ref{eq:epicdefJ}), inversely
proportional to the root mean square of the epicyclic
amplitude. This means that, for the same perturbing potential, a
given resonance will absorb, or emit, considerably more angular momentum
if its stars are cold, rather than hot. 

\subsection{Bulge and halo components}
\label{sec:halo_analytic}

Like the particles in the disc component, for a steady forcing, particles in the bulge
or the halo can emit/absorb angular momentum only if they are at
resonance. In the case of a growing or decaying perturbation, all
particles can emit/absorb angular momentum, but the amount is small
unless the perturbation is strongly growing or decaying. A02a
showed that the halo component can have a considerable amount
of resonant and near-resonant orbits, because of its response to the
bar. The distribution functions of the halo and bulge should be less
sharply peaked
than that of the disc because these components are hotter. Therefore
one expects the amount of angular momentum 
emitted or absorbed per unit mass at resonance to be smaller for the
halo than for the disc component. But, since the halo is 
heavier than the disc, its contribution to the total angular momentum
exchange could be considerable. 

The epicyclic approximation can not be applied to the
spheroids. Eq.~(\ref{eq:DLgeneral2}) can, nevertheless, be simplified
by assuming that the distribution function
depends only on the energy. Using eq.~(\ref{eq:motion}), I get

\begin{equation}
l\frac{\partial F}{\partial J_1}+m\frac{\partial F}{\partial J_2}
+n\frac{\partial F}{\partial J_3} = \frac{\partial F}{\partial E}~
(l\Omega_1 + m\Omega_2 + n\Omega_3).
\end{equation}

\noindent
Then, using the resonant condition (\ref{eq:resonances}),
eq.~(\ref{eq:DLgeneral2}) can be written  

\begin{eqnarray} 
\dot L_z = - \frac {1}{8\pi} 
      \int \int \int dJ_1 dJ_2 dJ_3~~~~~~~~~~~~~~~~~~~~~~~~~~~~~~~
      \nonumber \\
      \times~~\sum_{l,m,n}\
       m^2 \Omega_p \left(\frac{\partial F}{\partial E}
\right) |\psi_{lmn}|^2 
\delta (l\Omega_1+m\Omega_2+n\Omega_3+\omega).
\label{eq:DLgeneralh2}
\end{eqnarray}

\noindent
Note that all terms in the summation have the same sign,
independent of the values of $l, m$ and $n$. In the physically
reasonable case where the distribution function is a decreasing
function of the energy, I get $\dot L_z > 0$, which means that, as long as
the distribution function depends only on the energy, {\it all}
halo or bulge resonances gain angular momentum. 

Similarly eq. (\ref{eq:DLgeneral}) simplifies to

\begin{eqnarray} 
\Delta L_z = - \frac {1}{16\pi^2} e^{-2\omega_It}
      \int \int \int dJ_1 dJ_2 dJ_3~~~~~~~~~~~~~~~~~ 
      \nonumber \\
      \times~~\sum_{l,m,n}\
      \frac{m^2~\Omega_p~\frac{\partial F}{\partial E}~|\psi_{lmn}|^2}
{|l\Omega_1+m\Omega_2+n\Omega_3+\omega|^2}. 
\label{eq:DLgeneralh}
\end{eqnarray}

\noindent
This gives the total angular momentum that will be gained by the
spheroid. Since $\frac{\partial F}{\partial E}$ is absolutely larger for
colder distributions, these will be able to absorb more angular momentum
than hotter ones. 

\section{Resonances and orbital structure in the disc component}
\label{sec:orbits}

The previous section underlined the importance of resonances in the
evolution of the galaxy. Here I will discuss the orbital structure at
resonances, focusing on the parts that will be essential in
understanding the simulation results described in the following
sections.

\subsection{Resonant orbits}

The planar resonances occur for $n$ = 0. With this restriction and using
eqs.~(\ref{eq:epicdefO}), the resonant condition ~(\ref{eq:resonances})
can be simply written as  

\begin{equation}
l\kappa+m\Omega = - \omega_R = m \Omega_p
\label{eq:resondef} 
\end{equation}

\noindent
For $l$ = --1 and $m$ = 2 one has the inner Lindblad resonance
(hereafter ILR), while for $l$ = 1 and $m$ = 2 one has the outer
Lindblad resonance (hereafter OLR). For the former $lm < 0$, so that
disc particles 
at this resonance will lose angular momentum, the opposite being true
for the OLR. The same will be true for larger values of $m$, so that
particles at the $(l, m)$ = (--1, 3), (-1, 4), (-1, 5) etc. resonances will lose
angular momentum and particles at the $(l, m)$ = (1, 3), (1, 4), (1, 5) etc.
resonances will gain it. There are also resonances for $|l| > 1$, but
these are higher order and therefore should be of lesser dynamical
importance. For $l$ = 0 one has the corotation radius (hereafter CR),
at which the angular frequency of the particle is equal to the pattern
frequency. 

Resonant orbits are easy to visualise. For example
orbits at the ILR will, in the frame of reference of the bar, close
after one revolution around the center 
and two radial oscillations. Similarly, orbits at the (-1,~4) resonance
will close after one revolution and four radial oscillations
etc. Orbits at the higher order resonances, where $|l| > 1$, will close
after more than one revolution.

Several papers have focused on the study of periodic orbits in barred
galaxy potentials (see Contopoulos \& Grosb{\o}l 1989, for a
review). They discuss in detail the properties of the $x_1$ and the
$x_1$-related orbits, which are the backbone of all orbital structure 
(Athanassoula, Bienayme, Martinet
et al. 1983, Skokos, Patsis \& Athanassoula 2002). The $x_1$ are
periodic orbits 
that close after one revolution and two radial oscillations. {\it Therefore
all $x_1$ orbits are $l$ = -1 and $m$ = 2 resonant orbits, i.e. ILR
resonant orbits}. This
simple fact has not been generally noted so far, and this has led to a
number of misunderstandings, one of which I will discuss below, since
it is linked with the very definition of the ILR.
The $x_1$-related orbits are 3D periodic orbits whose families
bifurcate from the vertical instabilities of the $x_1$ family. Their
properties have been described by Skokos et al. (2002). At higher
energies and farther from the center one finds the $(1, 3)$, $(1, 4)$,
$(1, 5)$ etc. families of periodic orbits. Their 
orbits close after one revolution and 3, 4, 5, ... oscillations,
respectively. Members of these families are therefore also resonant
orbits with $l$ = -1 and $m$ = 3, 4, 5 etc.. As $m$ increases we
approach corotation, but families with high $m$ values are crowded
together on the characteristic diagram (e.g. fig. 2 of Athanassoula
1992) and have less extended stable parts. A similar sequence can be
found outside corotation for $l$ = 1, starting with large $m$ and
moving outwards to the OLR as $m$ decreases. Periodic orbits of both
the outer and the inner sequence of families have been calculated for
many barred potentials (see e.g. Contopoulos \& Grosb{\o}l 1989 for a
review). Such
orbital studies have also used surfaces of section to show that 
stable members of all these families can trap around them other
orbits, which, in orbital structure studies, are usually referred
to as regular or trapped orbits. They are in fact 
near-resonant orbits and can be thought of as a superposition of a
periodic/resonant orbit and an oscillation around it.     
It is therefore possible to use all the results on the structure and
stability of periodic orbits obtained in the extensive literature on
the subject to the study of resonances and vice-versa. Unfortunately
the terminology in these two fields is not always in agreement. A
clear case of disagreement concerns the very definition of the ILR and
its orbits, and stems from the fact that the frequencies $\Omega$ and
$\kappa$ of a single particle are equal to those of the galaxy at that
radius only if the galaxy is near-axisymmetric and cold. 

\subsection{Definition of the ILR}

Let us consider an axisymmetric galaxy with a non-axisymmetric
small perturbation of pattern speed $\Omega_p$. According to the
standard linear definition, an ILR occurs if and where

\begin{equation}
\Omega - \kappa / 2 = \Omega_p. 
\label{eq:ILR}
\end{equation}

\noindent
This may happen at one, two or no
radii, which are called the ILR radii, and one correspondingly says the
galaxy has one, two or no ILRs. 
Of course this definition is valid only for galaxies
with non-axisymmetric perturbations of very low amplitude (e.g. Binney
and Tremaine 1987). It can not be applied directly to strongly
non-axisymmetric galaxies or models, since for such cases the
potential and forces are a function of the angle as well as the
radius, and thus $\Omega$ and $\kappa$ can not be strictly defined and used as
in the linear case. It is thus necessary to consider some extension of
the linear definition. More than one is possible, and at least three
have been so far proposed, which may in some cases contradict each other.

One way of extending the linear definition is by noting that the
angular and epicyclic frequencies are 
the natural frequencies of the orbits in a given potential. Then the
linear definition of the ILR is extended to eq.~(\ref{eq:resondef}).
Strictly speaking, this is the definition of a resonant orbit, not the
definition of the existence of the resonance or of its radius. But one
can extend it to say that a galaxy has an ILR resonance if it has ILR
resonant orbits.  

A different extension was adopted in orbital structure studies. 
In the axisymmetric -- or near-axisymmetric -- cases the various $(l, m)$
families bifurcate from the $x_1$ at the corresponding resonances. This can be
extended to the strongly nonlinear and non-axisymmetric cases by saying
that resonances occur
where there are gaps or bifurcations introducing new $(l, m)$
families. In particular, in the axisymmetric case the families of $l$
= -1, $m$ = 2 orbits are bifurcated at the $(-1, 2)$ resonance(s).
These families are called $x_2$ and $x_3$ and their orbits are oriented perpendicular to the
bar. As the amplitude of the perturbation increases the bifurcations 
are substituted by gaps, but one can still say that an ILR resonance exists if and only
if these $x_2$ and $x_3$ families exist (e.g. van Albada \& Sanders 1982,
Athanassoula 1992).  

Finally in observational work, and often in the analysis of
$N$-body simulations, the linear axisymmetric definition is 
extended very simply by defining for a given quantity, e.g. the mass,
its axisymmetric equivalent, simply by averaging this quantity over
the azimuthal angle. Thus the linear definition can be applied to
strongly nonlinear and non-axisymmetric cases (e.g. 
Sanders \& Tubbs 1980, Reynaud \& Downes 1997).

Each of these extensions and corresponding definitions is reasonable
and useful within its own context. They are, however, not fully
compatible, and, in many cases, can lead to contradictions. For
example with the first 
extension all barred galaxies have ILRs, since all have $x_1$ orbits,
i.e. $l$ = -1, $m$
= 2 resonant orbits, and the resonant region is very broad,
comparable to the bar size. The same barred galaxies, however, may, by
the orbital structure definition, not have ILRs, if their potentials do
not allow $x_2$ and $x_3$ families. Similarly Athanassoula (1992)
showed a few examples where the orbital structure definition and the observational
definition are in contradiction (although in general there is
agreement). It is beyond the scope of this paper 
to propose a solution to this nomenclature problem. I, nevertheless,
want to underline it here, since it has often led to misunderstandings.  

\subsection{Orbits and bar angular momentum}

Viewing the bar as an ensemble of orbits allows considerable insight.
It shows that the bar has several, dynamically connected, ways of losing angular
momentum. First, by trapping particles which were
on quasi-circular orbit outside the bar, into elongated orbits in its
outer parts. This way angular momentum is lost from the disc inner
parts, while the bar will become longer, i.e. stronger. A second
alternative is if part, or all, of the orbits trapped in the bar become
more elongated. This way the bar looses angular momentum, while
becoming thinner, i.e. stronger. Finally the bar can slow down its
figure rotation, i.e. decrease its pattern speed, and again lose
angular momentum. In fact these three possibilities should be
linked. For example if 
a bar becomes longer, then it might have also to slow down in order to
push its CR further out and thus make space for the newly trapped
orbits in its outer parts (which must necessarily lie within CR). In
section \ref{sec:correlations} I will also show that there is a general
anti-correlation between the strength and the pattern speed of a
bar. This shows that bars use more than one of the three
alternatives at their disposal. Analytical calculations, however, are
not capable of determining the extent to which each of the three
alternatives will be used in a specific case. 

The linear theory described in the first part of
section~\ref{sec:analytic} clearly predicts that more angular
momentum can be emitted or absorbed at a given resonance if there are
more absorbers/emitters and/or if they are colder. The change of the
pattern speed with time enters only via the nonlinear equations
(\ref{eq:Lomp}) and (\ref{eq:Ldomp}), which relate the angular momentum of the bar
with its strength and pattern speed. The change of pattern speed
effects strongly the equilibrium between emitters and absorbers, since,
to a first approximation, the two are divided by the CR. 
Since I here consider isolated galaxies, the total amount of angular
momentum emitted at any time should be equal to the total amount
absorbed. For lower values of the pattern speed corotation  
will be further out, so that there will be more particles trapped in
the inner, $(-1, m)$, resonances which emit angular momentum. On the
other hand, the resonances which absorb angular momentum will also be
further out, both for the disc and the halo, and thus in regions of
lower density, where less material can be marshaled into absorbing
angular momentum. Thus lowering the pattern speed favours emitters
and disfavours absorbers. There should thus be, for every
case, an optimum 
radius dividing emitters from absorbers, for which emission will
balance absorption. This optimum
radius is linked to corotation and therefore to the pattern speed. 

Pushed further, this line of thought allows us to predict which
configurations will favour faster pattern speed and which slower
ones. Indeed, in cases where the halo can not absorb much angular
momentum -- either because it has low density in the relevant regions,
or because it is very hot -- the role of the outer disc is
important. Thus CR should not be too far out, so as to leave
sufficient space for disc absorbers. The opposite can be the case for
models, or galaxies, where the haloes can absorb considerable amounts
of angular momentum. In principle, and provided the halo is
sufficiently receptive, CR could be located in the outermost parts of
the disc.

\section{Simulations and numerical miscelanea}
\label{sec:simulations}

$N$-body simulations are a much easier test bed of the above theoretical
predictions than real galaxies, since they allow us to `observe' the
halo component and also to `observe' time evolution.
In the remaining sections I will use them for this purpose.  

The galaxies I model numerically consist initially of a disc, a halo, and sometimes a bulge
component. The density distribution in the disc is given by

\begin{equation}
\rho_d (R, z) = \frac {M_d}{4 \pi R_d^2 z_0}~~\exp (- R/R_d)~~{\rm sech}^2 (z/z_0),
\end{equation}

\noindent
that in the bulge by

\begin{equation}
\rho_b (r) = \frac {M_b}{2\pi a^2}~~\frac {1}{r(1+r/a)^3},
\end{equation}

\noindent
and that in the halo by

\begin{equation}
\rho_h (r) = \frac {M_h}{2\pi^{3/2}}~~ \frac{\alpha}{r_c} ~~\frac {\exp(-r^2/r_c^
2)}{r^2+\gamma^2}.
\label{eq:halodens}
\end{equation}

\noindent
In the above $r$ is the radius, $R$ is the cylindrical radius, $M_d$,
$M_b$ and $M_h$ are the masses of the disc, bulge 
and halo respectively, $R_d$ is the disc radial scale length, $z_0$ is
the disc vertical scale thickness, $a$ is the bulge scale length,
and $\gamma$ and $r_c$ are halo scale lengths. The former can be
considered as a core radius, and will hereafter be loosely referred to
as such.
The parameter $\alpha$ in the halo density equation is a normalisation
constant defined by 

\begin{equation}
\alpha = [1 - \sqrt \pi~q~\exp (q^2)~~(1 - {\rm erf} (q))]^{-1},
\end{equation} 

\noindent
where $q=\gamma / r_c$ (Hernquist 1993). The halo velocity 
distribution is isotropic. In building the initial conditions
I loosely followed Hernquist (1993) and Athanassoula \& Misiriotis 
(2002, hereafter AM02). 
In some cases, as e.g. those
described in section~\ref{sec:halodisp}, I need a considerably more
extended halo. I then use for the density the sum of two densities as
in (\ref{eq:halodens}), making sure
that the two put together give a reasonable total halo. When
describing such cases, I will use the subscript 1 for all quantities
referring to the inner halo component and 2 for those of the outer
halo. The functional forms for the two are identical.

\begin{table*}
\caption[]{Initial conditions for some of the models}
\begin{flushleft}
\label{tab:initcond}
\begin{tabular}{lllllllllllllll}
\hline
model & $M_d$ & $R_d$ & $z_0$ & $Q$ & $Q(r)$ & $M_{b}$ & $a$ & $M_{h1}$ &
$\gamma_1$ & $r_{c1}$ & $M_{h2}$ & $\gamma_2$ & $r_{c2}$ &
$r_{trunc}$ \\
\hline
MB0 & 1. & 1. & 0.1 & 0.9 & c & 0.6 & 0.4 & 5. & 6. & 10. & 0 & -- &
-- & 15 \\
MQ1 & 1. & 1. & 0.2 & 0.1 & c & 0 & -- & 5. & 0.5 & 10. & 0 & -- & -- & 15  \\
MQ2 & 1. & 1. & 0.2 & 0.9 & c & 0 & -- & 5. & 0.5 & 10. & 0 & -- & -- & 15 \\
MQ3 & 1. & 1. & 0.2 & 1.2 & c & 0 & -- & 5. & 0.5 & 10. & 0 & -- & -- & 15 \\
MQ4 & 1. & 1. & 0.2 & 1.4 & c & 0 & -- & 5. & 0.5 & 10. & 0 & -- & -- & 15 \\
MQ5 & 1. & 1. & 0.2 & 1.6 & c & 0 & -- & 5. & 0.5 & 10. & 0 & -- & -- & 15 \\
MQ6 & 1. & 1. & 0.2 & 1.8 & c & 0 & -- & 5. & 0.5 & 10. & 0 & -- & -- & 15 \\
MQ7 & 1. & 1. & 0.2 & 2.0 & c & 0 & -- & 5. & 0.5 & 10. & 0 & -- & -- & 15 \\
MQ8 & 1. & 1. & 0.2 & 2.2 & c & 0 & -- & 5. & 0.5 & 10. & 0 & -- & -- & 15 \\
MQV1 & 1. & 1. & 0.2 & 0.9 & v & 0 & -- & 5. & 0.5 & 10. & 0 & -- &
-- & 15\\
MQV2 & 1. & 1. & 0.2 & 1. & v & 0 & -- & 5. & 5. & 10. & 0 & -- &
-- & 15  \\
MDQ & 1. & 1. & 0.2 & 1. & c & 0 & -- & 5. & 5. & 10. & 0 & -- & -- & 15 \\
M$\gamma1$ & 1. & 1. & 0.2 & 1.2 & c & 0 & -- & 5. & 0.01 & 10. & 0 & -- & -- & 15 \\
M$\gamma2$ & 1. & 1. & 0.2 & 1.2 & c & 0 & -- & 5. & 0.1 & 10. & 0 & -- & -- & 15 \\
M$\gamma3$ & 1. & 1. & 0.2 & 1.2 & c & 0 & -- & 5. & 0.5 & 10. & 0 & -- & -- & 15 \\
M$\gamma4$ & 1. & 1. & 0.2 & 1.2 & c & 0 & -- & 5. & 1. & 10. & 0 & -- & -- & 15 \\
M$\gamma5$ & 1. & 1. & 0.2 & 1.2 & c & 0 & -- & 5. & 2.5 & 10. & 0 & -- & -- & 15 \\
M$\gamma6$ & 1. & 1. & 0.2 & 1.2 & c & 0 & -- & 5. & 4. & 10. & 0 & -- & -- & 15 \\
M$\gamma7$ & 1. & 1. & 0.2 & 1.2 & c & 0 & -- & 5. & 5. & 10. & 0 & -- & -- & 15 \\
MHH1 & 1. & 1. & 0.2 & 0.9 & c & 0 & -- & 5. & 0.5 & 10. & 10 & 10 &
15 & 25 \\
MHH2 & 1. & 1. & 0.2 & 0.9 & c & 0 & -- & 5. & 0.5 & 10. & 10 & 4 &
15 & 25 \\ 
MHH3 & 1. & 1. & 0.2 & 0.9 & c & 0 & -- & 5. & 0.5 & 10. & 20 & 10 &
15 & 25 \\ 
MH1 & 1. & 1. & 0.2 & 1.2 & c & 0 & -- & 2. & 0.5 & 10. & 0 & -- & -- & 15 \\
MH2 & 1. & 1. & 0.2 & 1.2 & c & 0 & -- & 1. & 0.5 & 10. & 0 & -- & -- & 15  \\
MH3 & 1. & 1. & 0.2 & 0.9 & c & 0 & -- & 6.25 & 0.5 & 10. & 0 & -- & -- & 15 \\
MH4 & 1. & 1. & 0.2 & 0.9 & c & 0 & -- & 8.333 & 0.5 & 10. & 0 & -- & -- & 15 \\
\hline
\end{tabular}
\end{flushleft}
\end{table*}

The results presented in this paper are based on 160 
simulations. These include mainly (roughly 80\%) simulations made with the
Marseille Grape-5 computers (for a description of Grape-5 boards and
their performance see Kawai et al. 2000), using a tree-code specifically  
adapted to this hardware, while roughly 20\% of the simulations were
run with W. Dehnen's treecode on PC workstations (Dehnen 2000, 2002). The
results from the two codes agree well, as I could show by running
a few simulations with both codes.
In most of the simulations all particles had the same mass, but in 10
simulations I used particles of different masses. To make sure that
this did not introduce any bias I first calculated the pericenter of
each particle orbit in the initial configuration and then sorted the
particles as a function of this quantity. The largest masses were then
attributed to particles with the largest pericenters, thus ensuring
that these particles will never reach too near the center. Since the halo
density distribution does not change much with time (AM02), these
particles will not 
visit the inner regions even after the model has evolved. Using the
pericenter in order to 
attribute the particle masses is much safer than using the
distance of the particle from the center at $t$ = 0, since many
particles will be near their apocenters at this time. Such particles,
following their orbits, may visit at later times the inner parts where
the disc with the lighter particles resides. This 
may introduce a bias in the evolution. 

In previous papers (e.g. AM02, A02 etc.) I called MH (for Massive
Halo) simulations with small cores, i.e. small $\gamma$ values, for
which the halo dominates 
in the inner parts. I called MD (for Massive Disc) simulations with
large cores ($\gamma$ values), where the disc dominates in the inner
parts. This will carry over in this paper, but since I will here discuss a
very large number of such simulations, I will rather use the names MH-type
and MD-type.

Listing the initial condition parameters for all 160 simulations would
be too lengthy, so I include in Table \ref{tab:initcond} only those
runs which will be discussed in more detail in this paper. From left  
to right the columns give the name of the run, the mass and scale length
of the exponential disc, the value of $Q$ at $R$ = 0.1 and a symbol 
denoting whether $Q$ is a function of radius (v), or not (c). Then 
follow the three parameters describing the halo, namely its mass and its two 
scale lengths. The next three columns give the same parameters for the
second halo, whenever this exists, and the last column gives
the truncation radius of the mass distribution. This
table lists all masses taken out to infinity.  

A reasonable calibration, allowing to convert computer units to
astronomical ones, has been given by AM02.
They have taken the unit of mass equal to 5 $\times$ $10^{10}$ M$_\odot$,
the unit of length equal to 3.5 kpc and $G$ = 1. This gives that the
unit of velocity is 248 km/sec and the unit of time is 1.4 $\times$
$10^7$ yrs. 
This calibration is not unique and other ones can be equally good,
depending on the application. I will thus present here all results in
computer units and let the reader do the conversion according to
his/her needs.
Care, however, is necessary if one wants to compare the results of the
simulations quantitatively to observations, since the length scale of
the disc evolves with time (Valenzuela \& Klypin 2002). Thus, when
comparing with observations, one
should not compare the observed disc scale length with the disc scale
length of the initial conditions. One should instead fix a time at
which the simulations should be compared to the observations, well
after the bar has grown, and then compare the scale length obtained
from that time to the observed one. Although this warning is in fact
just common sense, its 
neglect has led to a number of errors and misunderstandings. 

In most simulations the total number of particles used
was between 1 and 2 million and I used a softening of 0.0625, or, in some
cases, 0.03125. I repeated a few simulations 
with double and/or half the number of particles to make sure that the number used
was sufficient. I further assessed the numerical robustness of my
results by repeating some of the simulations with half and/or double the softening
length and/or time step. Finally the comparison of the results obtained
with the Grape-5 treecode, and those obtained with W. Dehnen's
treecode argue strongly for the reliability of the results. I made an
ultimate, very strong test, by running one case with a direct
summation code. Since this is the code that has the least numerical
assumptions, the very good agreement between those results and those
obtained with a Grape-5 treecode argues strongly for the
reliability of the code.

In simulations like those presented here one has to be particularly
careful that numerical problems do not stop the resonant particles from
gaining/loosing angular momentum. This would be the case if, for
example, the softening was too small for the adopted
number of particles, in which case the simulation would
be noisy. Encounters would be important and would knock particles
off their resonant trajectories and thus artificially limit the
angular momentum that could be emitted or absorbed by the
resonances. Therefore, a simulation with a smaller
softening is not necessarily better than one with a bigger softening. It
can in some cases be worse, if the number of particles in the disc
and/or halo is not sufficiently high. The most difficult cases to
check are those with individual softening and/or individual mass. Indeed,
it suffices that the 
simulation is noise dominated in one region for the resonant orbits
traversing that region to have a high probability of being
artificially scattered.

In this paper I discuss the bar strength and its evolution, and
compare the strength of bars in different simulations. Although the
notion of bar strength is clear to everyone, and it is very often easy,
when comparing two bars, to say which one is strongest, a precise
definition is not trivial. Several possibilities have been put forward
so far, ranging from the bar axial ratio, to some function of the
tangential forcing. Here I wish to stay as near as possible to the
theoretical work presented in section \ref{sec:analytic}, and so I will
use the $m$ = 2 component of the mass or density distribution in the
disc. This has the added advantage that it can also be applied to
observations, thus permitting a direct comparison between theory,
$N$-body simulations and observations. Thus, following the notation of AM02,
I quantify the bar strength with the help of the Fourier components
of the face-on density or mass distribution. For a given $m$ I will use the  
relative Fourier amplitudes $\sqrt{A_m^2+B_m^2}/ A_0$, where 
 
\begin{equation}
A_m (r) = \frac {1}{\pi} \int _0 ^{2 \pi} \Sigma (r, \theta) cos (m
\theta) d \theta, ~~~~~~~m=0, 1, 2, ...
\end{equation}

\begin{equation}
B_m (r) = \frac {1}{\pi} \int _0 ^{2 \pi} \Sigma (r, \theta) sin (m
\theta) d \theta, ~~~~~~~m=1, 2, ... 
\end{equation}

\noindent
and  $\Sigma (r, \theta)$ is the projected surface density, or mass. 
The above equations define the Fourier components, which are a function of the
radius. To measure the strength of the bar, however, I want a single
quantity, and not a function of 
radius. For this I take

\begin{equation}
S_B = \frac {\int _0 ^{R_{max}}\sqrt{A_m^2+B_m^2}~r~dr }  
{\int _0 ^{R_{max}} A_0~r~dr},
\end{equation}

\noindent
It is easy to extend this definition to include higher $m$ values, but
I have refrained from doing so in order to stay as near the
linear theory as possible. I have also used an alternative definition,
without the $r$ factor in the integrands of the numerator and
denominator. The results are qualitatively similar. 

\begin{figure*} 
\includegraphics{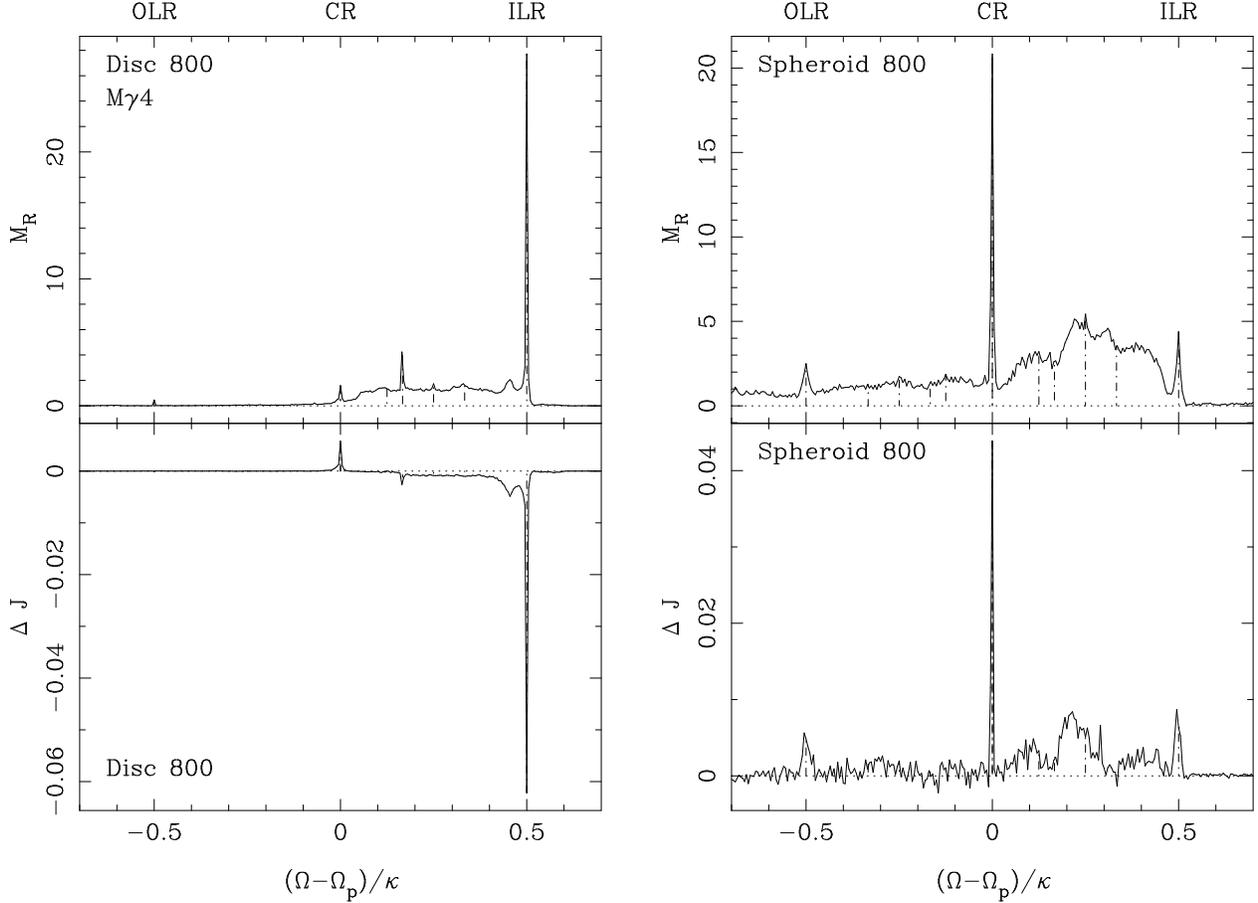} 
\vspace{12.cm}
\caption{The upper panels give, for time $t$ = 800, the mass per unit
frequency ratio, $M_R$, as a
function of that ratio. The lower panels give $\Delta J$, the angular momentum
gained or lost by the particles between times 800 and 500, plotted as
a function of their frequency ratio $(\Omega - \Omega_p) / \kappa$,
calculated at time $t$ = 800. The left panels correspond to the disc
component and the right ones to the spheroid. The
component and the time are written in the upper left corner of each
panel. The vertical dot-dashed lines give the positions of the main
resonances.  
}
\label{fig:resonances}
\end{figure*}

In section \ref{sec:angmom} I will discuss resonant stars and
the angular momentum they lose or gain. The procedure involved is a
straightforward extension of that 
outlined in A02. I first freeze the potential at a given time in
the simulation, allowing only the bar to rotate rigidly with a pattern
speed equal to that in the simulation. I then 
calculate in this potential the orbits of 100 000 particles 
taken at random from the disc population, and 100 000 particles taken
at random from the spheroid population. For each orbit, I calculate
the principal frequencies $\Omega$ and $\kappa$ using
spectral analysis. Since the angular frequency proved more difficult
to calculate reliably, I supplemented the spectral analysis with other,
more straightforward methods, based on following the angle as a
function of time. I then binned the particles in bins of width $\Delta
(\frac {\Omega - \Omega_p} {\kappa})$ = 0.005, calculated the mass
within each bin, and, dividing the mass in the
bin by its width, I obtained the mass per unit frequency of the bin,
$M_R$. This is somewhat different from what I did in A02, 
where I used number, rather than mass, densities. Number and mass
densities are equivalent for A02, since in the simulations discussed
there all particles have the same mass. Since now in a few
of my simulations this is not the case, it was
necessary to introduce this modification. I also calculated the mean angular momentum of the
orbit, where the time average was taken over 40 bar rotations of the
frozen potential. One can thus associate at any given time a frequency ratio
$(\Omega - \Omega_p) / \kappa$ and an angular momentum. 
The error estimates are as discussed in A02, and the orbits for which
the estimates of the frequencies were not considered sufficiently
reliable were not used in any further analysis. This was of the order
of, or less than, 15\%. 

\section{Angular momentum exchange}
\label{sec:angmom}

I will use the $N$-body simulations introduced above to test whether the
analytical results developed in section~\ref{sec:analytic} can still be
applied to cases as non-axisymmetric and as non-linear as strongly
barred galaxies. The linear theory predicts
that angular momentum is gained by the resonant particles in the halo, the
bulge and the outer disc and is lost by the resonant particles in the
bar. Since the bar is a strongly non-linear feature, the
agreement should be, at best, qualitative, rather than quantitative. 

\begin{figure} 
\includegraphics{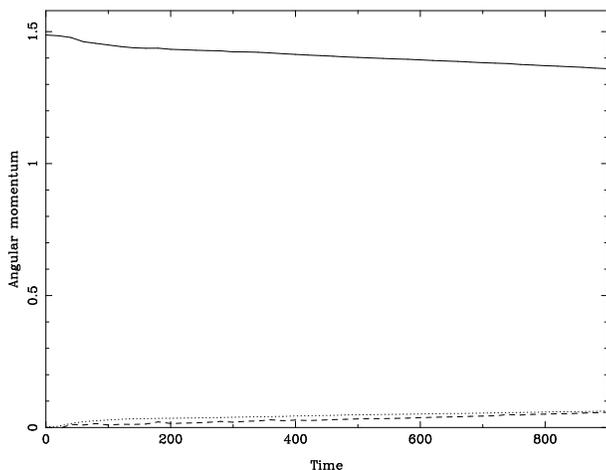} 
\vspace{7.cm}
\caption{Angular momentum as a function of time for the disc component
(solid line), the halo (dashed) and the bulge (dotted) of model MB0. 
}
\label{fig:tangmom}
\end{figure}

In A02 I showed that there are indeed a lot of resonant and near-resonant
particles, both in the disc and the halo. It is possible to extend this
work to test for the angular momentum lost or gained at various
resonances. The results, for a run listed in Table \ref{tab:initcond} as 
M$\gamma$4, are plotted in fig.~\ref{fig:resonances}. The upper panels show,
for time $t$ = 800, the mass per unit frequency ratio $M_R$ of particles
having a given value of the 
frequency ratio $(\Omega - \Omega_p) / \kappa$ as a function of this
frequency ratio. As was already shown in A02,
there are clear peaks, indicating the existence of resonances. The
highest peak for the disc component is at the ILR, followed by
$(-1, 6)$ and CR. In all simulations with strong bars 
the ILR peak is strong. The CR peak is also always present, but not always of
the same amplitude. For MH-type cases, as the one shown in
fig.~\ref{fig:resonances}, it is small, while in MD-types it is very
important. The existence of peaks at other resonances, as well as their
importance, varies from one run to
another and also during the evolution of a given run. For the
spheroidal component the highest peak is at CR, followed by peaks 
at the ILR and OLR. The peaks at the OLR are more important, both for
the disc and the spheroid, for MD-type simulations. 

The lower panels show the angular
momentum exchanged. For this I calculated the angular momentum of each
particle at time 800 and time 500, as described in
section~\ref{sec:simulations}, and plotted the difference as a
function of the frequency ratio of the particle at time
800. It is clear from the figure that disc particles at ILR lose angular
momentum, while those at CR gain it. The $(-1, 6)$ also loses angular
momentum. There is a also a general, albeit small, loss of angular
momentum from particles with frequencies between CR and ILR. This
seems more important for simulations with haloes with smaller cores, as
e.g. M$\gamma$1 and M$\gamma$2. It could be partly due to particles
trapped around secondary resonances, and partly due to angular
momentum taken from particles which are neither resonant, nor
near-resonant, but can still lose a small amount of angular momentum
because the bar is growing, as was discussed in section
\ref{sec:analytic}. The corresponding panel for
the spheroid is, as expected, more noisy, but shows that particles at all
resonances gain angular momentum. The gain at the OLR is more
prominent for MD-type simulations, both for the disc and the spheroid
component. To summarise we can say that this plot, and similar ones
which I did for other simulations, confirm the results of section
\ref{sec:analytic} and show that the linear results concerning the
angular momentum gain or loss by resonant particles, 
qualitatively at least, carry over to the strongly nonlinear regime.

The evolution of the total angular momentum of the three components,
disc, halo and bulge, for simulation MB0 is given in
fig.~\ref{fig:tangmom}. It shows what one would expect after having
seen fig.~\ref{fig:resonances}, namely that the angular momentum of
the halo and bulge components increases steadily with time, while that
of the disc decreases. I made such plots for all my simulations and
found qualitatively similar results. However, the amount of angular
momentum exchanged varied widely from one simulation to another,
varying from just a few percent, to near 35 percent. Linear theory
predicts that the angular momentum exchanged would be highest if the
disc and spheroid components have high density and low velocity
dispersion, at least in the regions of the resonances. In such cases
the bars should be strong and their pattern speed
would decrease strongly with time. Thus the wide possible range of
exchanged angular momentum should be followed by a wide range of
possible bar strengths and possible pattern speed decreases. I will
address this prediction in the following sections.

\section{Velocity dispersion in the disc component}
\label{sec:discdisp}

The disc mass and scale length are used here as units of mass and
length. Therefore the only disc property that I can alter in order to influence
the angular momentum exchange is its velocity dispersion.
The linear theory given in section~\ref{sec:analytic} predicts that for a
given bar strength, the 
amount of angular momentum that can be emitted or absorbed at a disc
resonance, depends on the velocity dispersion of the disc material. Thus a cold
disc should emit/absorb considerably more at a given resonance than a hot
one. Furthermore, from 
eq. (\ref{eq:Lomp}) one can see that this should influence the pattern
speed of the bar. 

I made two different types of simulations in order
to assess these effects.
In the first class of simulations I used initial conditions where
$Q_{init}$ is constant with radius (as in AM02), and compared the results for
different values of $Q_{init}$. Preliminary results of these
simulations can be found in Athanassoula (2002b), and I will discuss
them further here.

\begin{figure} 
\includegraphics{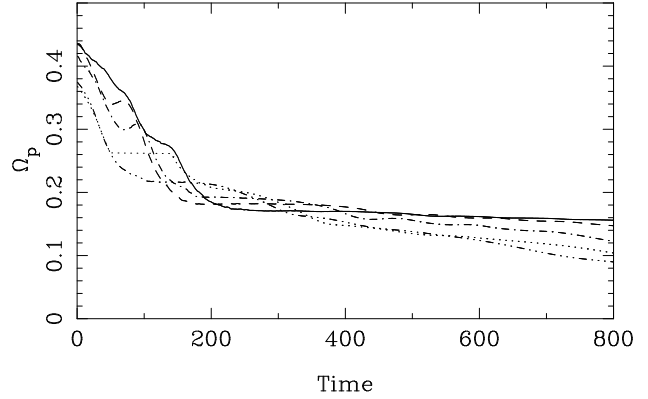} 
\vspace{6.cm}
\caption{Bar pattern speed as a function of time, for simulations
MQ4 ($Q_{init}$ = 1.4, dot-dot-dot-dashed), MQ5 (1.6, dotted),
MQ6 (1.8, dot-dashed), MQ7 (2, dashed) and MQ8 (2.2, solid line). 
}
\label{fig:Q1}
\end{figure}

\begin{figure} 
\includegraphics{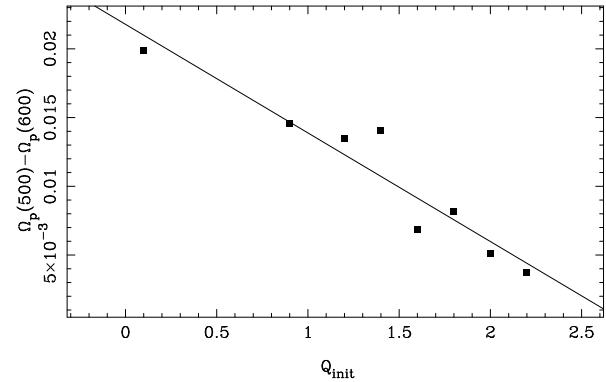} 
\vspace{6.cm}
\caption{Slowdown of the bar between times
500 and 600 as a function of $Q_{init}$ for simulations MQ1 to MQ8.
}
\label{fig:Q2}
\end{figure}

I ran a sequence of 8 simulations, with $M_d$ = 1, $M_h$ = 5, $R_d$ =
1, $z_0$ = 0.2, $\gamma$ = 0.5 and $r_c$ = 10. They differ by their
$Q_{init}$ value, which was for the coldest 0.1 and for the hottest
2.2 and are listed in Table \ref{tab:initcond} as MQ1 to MQ8. This
sequence of simulations shows clearly that the decrease of the bar
pattern speed  
with time is a function of the velocity dispersion of the disc
particles, in the sense that in colder discs the bar pattern speed
decreases much faster than in hotter discs. In the limit of a
sufficiently high $Q_{init}$, there is hardly any decrease. This is illustrated
in Fig.~\ref{fig:Q1}, where I plot the pattern speed as a function of
time for the five hottest simulations in the sequence, with $Q_{init}$
ranging between 1.4 and 2.2. During the initial stages of the
evolution the bar is not well developed, so its pattern speed is not
well determined. One should thus not heed the initial abrupt
decrease. After the bar has fully developed the pattern speed shows a
very strong decline with time for the coldest case
and hardly any decline for the hottest one. All 8 simulations are compared
in Fig.~\ref{fig:Q2}, where I plot the slowdown of the bar between
times 500 and 600 as a function of $Q_{init}$. It confirms the trend
seen in Fig.~\ref{fig:Q1}, showing a definite dependence of the
slowdown rate on $Q_{init}$. The coldest simulation, for $Q_{init}$ =
0.1, has well over four times the slowdown rate of the hottest
simulation, for $Q_{init}$ = 2.2. For the latter the slowdown rate is
indeed very small, less than 0.005, i.e. of the order of three percent in  
$\Delta t = 100$,
or, using the calibration proposed in section~\ref{sec:simulations},
in 1.4 Gyrs. In Fig.~\ref{fig:Q2} I have
also added the least squares fit to the data, which is only meant to
guide the eye. 

I have also made two simulations for which $Q_{init}$ decreases
exponentially with radius (Hernquist 1993). An MH-type such
simulation is listed in
Table~\ref{tab:initcond} as simulation MQV1. I have 
taken $Q_{init}(r = 0.1) = 0.9$, so that a comparison between MQV1
and MQ2 can show the effects of a hotter outer disc region. The
results of the two simulations differ only little. The difference between
the relative Fourier components is less than 10 per cent of the maximum and
is, after the bar has fully grown, always in the sense that run 
MQ2 has the strongest bar. The difference between the pattern speeds
is of the same order, or smaller, in the sense that for MQV1 the
pattern speed decreases less fast than MQ2. All this is in good
agreement with the picture 
of bar evolution presented here. Indeed in run MQ2 most of the
absorbing material is in the halo and only a small fraction is in the
disc. Thus, making this small fraction less efficient, by heating
the corresponding part of the disc, does not have much influence. In
the inner disc part, where the emitters are situated, the difference
in the initial $Q$ is not large. Thus the results of the two runs do
not differ much. The small difference between the two runs is also in
the sense that could have been predicted by the picture of bar
evolution presented here. Indeed MQ2 has the somewhat
stronger bar, whose pattern decreases somewhat faster, which could be
predicted since the outer parts of the disc are hotter.

An MD-type simulation with a variable $Q_{init}$ is listed in
Table~\ref{tab:initcond} as MQV2. In this case $Q_{init}(r = 0.1) =
1$, so that a comparison between MQV2 and MDQ can show the effects of
a hotter outer disc in MD-type simulations. This is much larger than
in the case of MH-types. Indeed the $m$ = 2 Fourier component drops by
30 to 40 percent, while the difference in the slowdown rate is
considerably larger. Again this is in good agreement with the picture
of bar evolution presented here. In MD-types the outer parts of the
disc should contribute significantly to the angular momentum absorption. Thus
making them hotter should lead to a considerably weaker bar whose
pattern speed decreases considerably less. This is indeed borne out by
the comparison of MQV2 and MDQ.
 
\section{Halo-to-disc mass ratio}
\label{sec:density}

The analysis in section~\ref{sec:analytic} shows that, all other
parameters being the same, the amount 
of angular momentum that can be emitted/absorbed at a given resonance is
proportional to the amount of mass at this resonance. In this section
I will test this with the help of $N$-body simulations with haloes 
of different mass, or of different mass distribution. 

\subsection{Halo density}
\label{sec:halodens}

\begin{figure} 
\includegraphics{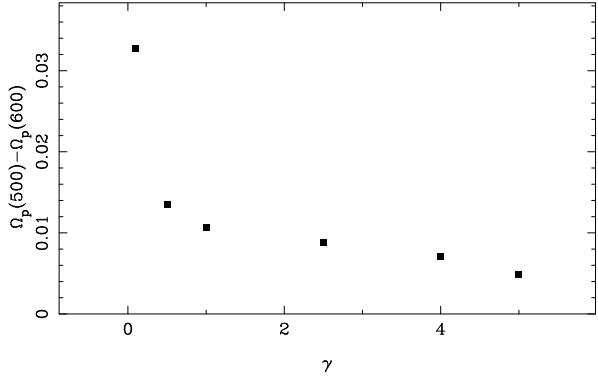} 
\vspace{6.cm}
\caption{Slowdown of the bar pattern speed between times
500 and 600 as a function of the halo central concentration for
simulations M$\gamma1$, M$\gamma2$, M$\gamma3$, M$\gamma4$,
M$\gamma5$, M$\gamma6$ and M$\gamma7$. 
 }
\label{fig:omp_gamma1}
\end{figure}

\begin{figure} 
\includegraphics{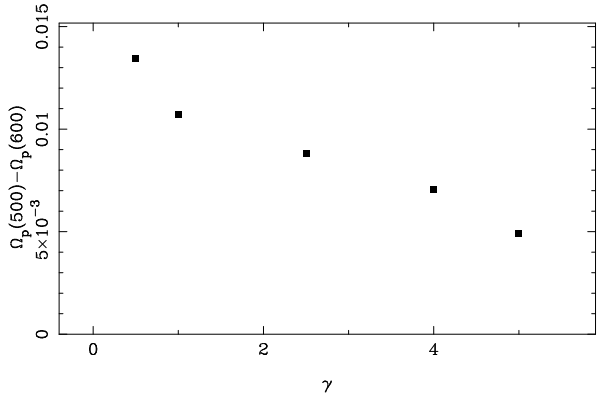} 
\vspace{6.cm}
\caption{Blow-up of the lower part of the previous figure, showing 
that the trend with central concentration is clear even for higher
values of $\gamma$. 
 }
\label{fig:omp_gamma2}
\end{figure}

\begin{figure} 
\includegraphics{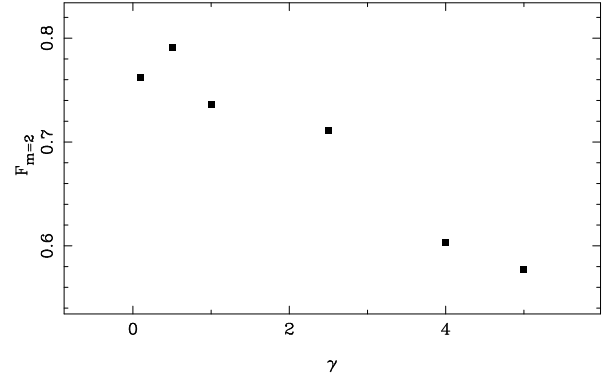} 
\vspace{6.cm}
\caption{Maximum of the relative $m$ = 2 Fourier component, $F_{m=2}$, at time 800
as a function of the initial halo central concentration for 
simulations M$\gamma1$, M$\gamma2$, M$\gamma3$, M$\gamma4$,
M$\gamma5$, M$\gamma6$ and M$\gamma7$. 
 }
\label{fig:strength_gamma}
\end{figure}

Some insight can be gained already by comparing
MH-type and MD-type models. Such comparisons, for various initial
condition parameters, can be found in AM02, A02, or Athanassoula
(2002b). These papers show clearly that stronger bars can grow in more
halo dominated models, so I do not include further figures here. Figures
\ref{fig:omp_gamma1} and \ref{fig:omp_gamma2} show the slowdown of the
bar pattern speed as a function of the halo core size 
$\gamma$ for a sequence of six models listed in Table
\ref{tab:initcond} as M$\gamma$2 to M$\gamma$7. As already mentioned,
MH-type haloes correspond to 
low values of $\gamma$ and MD-types to high values. As predicted,
MH-type models have pattern speeds that decrease faster than those of
MD-types and there is a definite trend between the slowdown rate and
$\gamma$. This is in agreement with the results of Debattista \&
Sellwood (1998, 2000), who find a strong temporal decrease of the
pattern speed for halo dominated cases and a weak one for disc
dominated cases. Figure \ref{fig:strength_gamma} plots $F_{m=2}$, the maximum of
the $m$ = 2 relative Fourier amplitude, as a function of the core
radius $\gamma$. It
shows a clear trend in the sense that the bars formed in more
concentrated haloes -- i.e. in MH-types -- are 
considerably stronger than those in less concentrated haloes --
i.e. in MD-types --. The trend is reversed in the innermost
parts. The angular
momentum gained by the halo also varies considerably along this
sequence of runs. In fact more angular momentum is exchanged in
simulations with lower $\gamma$, and that
up to and including $\gamma$ = 0.5. For this $\gamma$ value the angular
momentum gained by the halo is about 2.6 times higher than for $\gamma$
= 5. For the smallest value of $\gamma$ the angular momentum is
slightly smaller than for $\gamma$ = 0.5, i.e. show a similar behaviour
to that of the $F_{m=2}$. Thus this sequence of simulations shows
clearly that the strongest bars are associated with cases with more
angular momentum exchange.

\begin{figure*}
\includegraphics{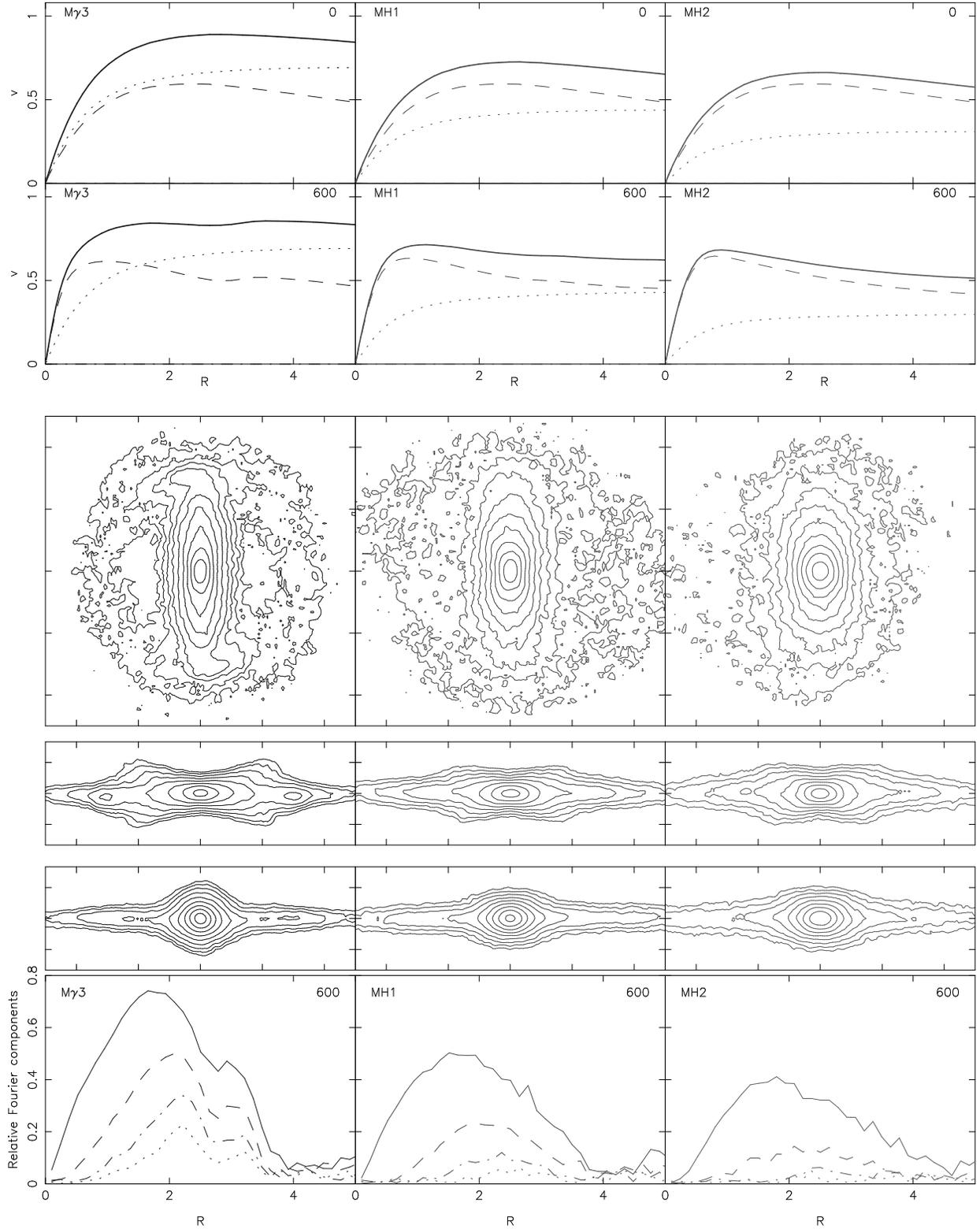}
\vspace{21.2cm}
\caption{Basic information on three simulations with different disc
  masses, for time $t$ = 600. Left
panels correspond to simulation M$\gamma3$, middle ones to simulation MH1 and
right ones to simulation MH2. The two upper rows give the circular
velocity curves at time 0 and 600. The dashed and dotted
lines give the contributions of the disc and halo
respectively, while the thick full lines give the total circular
velocity curves. The third row of panels gives the isocontours of the
density of the disc particles projected face-on and the fourth and
fifth row give the side-on and end-on edge-on views, respectively.
The side of the box for the face-on views is 10 length units and the height
of the box for the edge-on views is 3.33. The isodensities in the third row of
panels have been chosen so as to show best the
features in the bar and in the inner disc. No isodensities for the
outer disc have been included, although the disc extends well beyond the
area shown in the figure. The sixth row of panels gives the
$m$ = 2, 4, 6, and 8 Fourier components of the mass, as defined in AM02. 
}
\label{fig:basicMh}
\end{figure*}

\begin{figure} 
\includegraphics{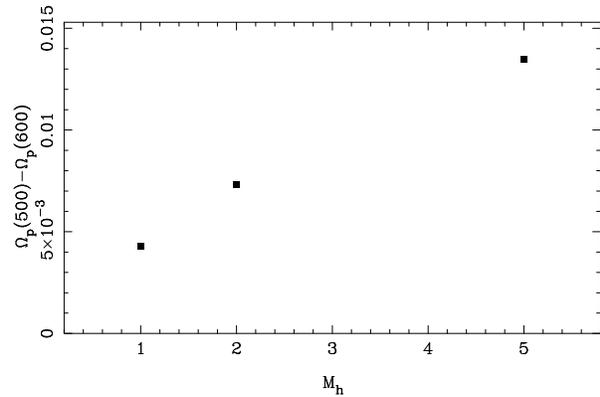} 
\vspace{6.cm}
\caption{Slowdown of the bar pattern speed between times
500 and 600 as a function of the halo mass for
simulations M$\gamma3$, MH1 and MH2. 
 }
\label{fig:omp_Mh}
\end{figure}

In order to understand better the effect of the halo on the evolution I will
now consider haloes with masses which are lower than those in the
simulations discussed so far. For this I will use 
a sequence of three simulations, all with $\gamma$ =
0.5 and different values of $M_h$. They are listed in Table
\ref{tab:initcond} as simulations M$\gamma3$, MH1 and MH2. Some of
their basic results are summarised in
Fig. \ref{fig:basicMh}. In the initial stages of the simulation the bar grows
slower in the more halo dominated environment, as expected
(e.g. Athanassoula \& Sellwood 1986). Nevertheless, the situation is
eventually reversed, so that at later times 
the simulation with the most massive halo has the
strongest bar. The other two simulations,
with considerably less massive haloes, have also less strong bars.
In particular, model MH2, which has the less massive halo, has
also the weakest bar, and its $m$ = 6 and 8 Fourier components do not
stand out clearly from the noise, as was the case for model MD in
AM02. The results of the three simulations differ also when seen
edge-on. Model MQ2 shows  
a very strong peanut when seen side-on (i.e. with the line of sight
along the bar minor axis), and a bulge-like protuberance out of the
equatorial plane when seen end-on (i.e. with the line of sight along
the bar major axis). As explained in AM02, this is just the bar seen
end-on, and not a real bulge. Models MH1 and MH2 viewed side-on
display a considerably less strong peanut, nearer to a boxy shape.
This sequence, the one shown in Fig.~\ref{fig:basicHH} and other similar 
ones not shown here, reinforce the result found by AM02, namely
that it is the strongest bars that form the strongest peanuts, or even
`X'-shapes, while milder bars form boxy shapes. This is in agreement
with observations (L\"utticke et al. 2000), although of course the
strength of the bar in edge-on galaxies can only be measured
indirectly from the density drops on cuts along the equatorial plane.

The decrease of the pattern speed with time is shown for these
three models in Fig. \ref{fig:omp_Mh}. There is a clear trend between
the decrease of pattern speed and the halo mass, in the sense that the 
heaviest halo mass has the fastest decreasing pattern speed. The
angular momentum gained by the halo also increases with the halo
mass. Thus the halo of model M$\gamma$3 gains somewhat more than twice
the angular momentum gained by the halo of model MH1, which in turn
gains somewhat more than twice that gained by MH2. A plot similar to
that of Fig. \ref{fig:resonances} shows that the height of the halo resonant peaks
decreases drastically as $M_h$ decreases. Thus this series of
simulations confirms beautifully the predictions of the bar evolution
picture presented here.   

A similar sequence of simulations, now for $\gamma$ = 5 (not listed in Table
\ref{tab:initcond}), shows different results (not plotted
here). Indeed now the strength of the bar depends little on $M_h$,
even though I have considered $M_h$ values ranging from 1 to 5.
The reason is that in MD cases, contrary to MH
ones, the outer disc has a considerable contributions to the angular
momentum absorption and 
is still capable of stepping in and providing the necessary sinks of
angular momentum when the halo contribution is low.   

The above should not give the false impression that by considering yet
more massive haloes we will always get yet stronger bars, whose pattern speed
will decrease yet faster. There is a limit beyond which the disc
self-gravity is so small that the bar can not grow. Furthermore, by
increasing the halo mass we also increase its velocity dispersion,
thus decreasing its responsiveness. This effect will be discussed in
detail in section~\ref{sec:halodisp}.

To check the above I ran a series of six
simulations with $M_h$ between 5 and 10. Three are are listed in
Table \ref{tab:initcond} as MQ2, MH3 and MH4. The initial 
growth rate of the bar is lower in simulations with a higher halo mass, in good
agreement with previous results (e.g. Athanassoula \& Sellwood 1986).
For $M_d/M_h$ = 0.2 the bar is very strong, while for 
$M_d/M_h$ = 0.16 both its strength and length are considerably smaller. In
fact the difference is rather strong, for a decrease of the
disc-to-halo mass ratio of only 20 per cent. A further 20 percent
decrease results in a density distribution which has only a mild
non-axisymmetry in the central parts, visible in the projected
isodensities and in the $m$ = 2 Fourier component. 

\begin{figure} 
\includegraphics{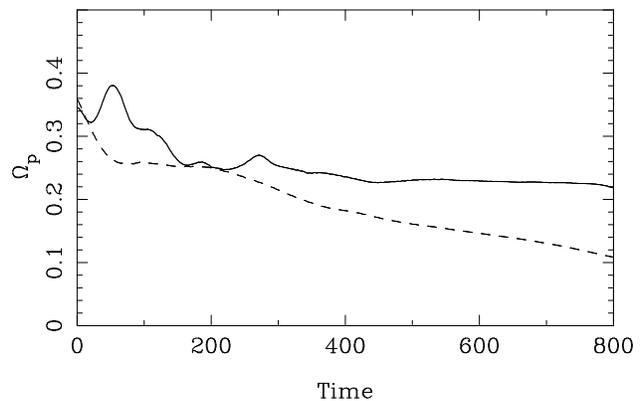} 
\vspace{6.cm}
\caption{Bar pattern speed as a function of time, for two simulations
with different disc-to-halo mass ratio. Simulation MQ2 (dashed line) has a
$M_d/M_h$ = 0.2, while simulation MH3 (solid line) has $M_d/M_h$ = 0.16. 
}
\label{fig:ompLSB}
\end{figure}

Fig. \ref{fig:ompLSB} shows the bar pattern speed as a function of
time, for two of the simulations discussed above. In simulation MQ2,
which has a substantial disc component, the pattern speed drops
considerably with time. On the other hand, the pattern speed hardly
decreases in simulation MH3, whose disc-to-halo mass ratio is only 20
per cent less massive. The amount of angular momentum gained
by the halo by 
time 900 is in MH3 5.7 times less than in MQ2. The difference is even
more extreme for simulation MH4, whose halo gains only three percent
of what was gained by the halo of MQ2. Thus this series of simulations
confirms the prediction that less angular momentum 
exchange within the galaxy results in 
a weaker bar, whose pattern speed decreases less fast.

\subsection{Bulge component}
\label{sec:bulge}

I also ran 13 simulations including a bulge component with a mass
between 0.2 and 0.6 and a scale length between
0.2 and 0.6. For MD type simulations the effect of the bulge is
quite pronounced, in the sense that the
strongest bars and peanuts form in models with the heaviest
bulges. One example of such a case can be seen in Figures 1 and 2 of
AM02, where one can compare the bar/peanut formed in simulations with
and without bulge (models MD and MDB of that paper). The effect of the bulge 
is not as noticeable in MH type simulations. 

\begin{figure} 
\includegraphics{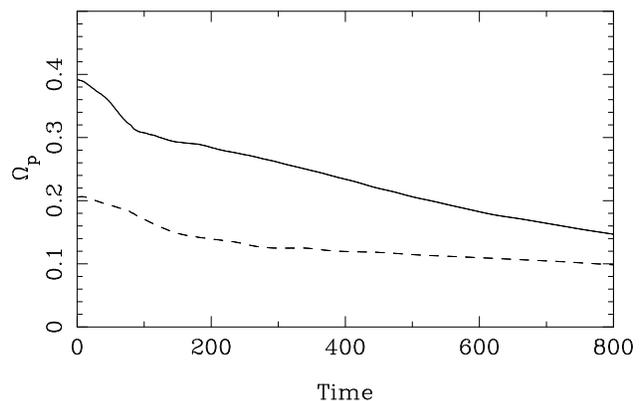} 
\vspace{6.cm}
\caption{Bar pattern speed as a function of time, for a simulation
with a bulge
($MDB$, solid line) and a simulation without (M$\gamma7$, dashed line). 
}
\label{fig:omp_bulge}
\end{figure}

The presence of the bulge also influences the slowdown rate of the
bar, in the sense that the pattern speed of bars decreases much faster
in the presence of a bulge than without it. The effect, as can be seen
from Fig. \ref{fig:omp_bulge}, can be quite strong. 

The effect of bulges, described above, is easily understood in the
framework of evolution via angular momentum exchange by resonant
stars. I have analysed the frequency of the orbits in the
spheroids (i.e. bulge and halo together), as in A02, and found
considerably more particles at resonance in cases with strong bulges.
There is also, in general, more angular momentum exchange. 
This can explain why such models have stronger bars and faster
slowdown rates.
 
Amongst the models I ran I have not encountered cases where a more
massive bulge resulted in a less strong bar, as was the case for
haloes, where, beyond a certain threshold, haloes hindered bar
formation. This, however, could be due to the fact that I did not
consider particularly massive bulges, or did not enhance their central
concentration sufficiently.

\section{Velocity dispersion of the halo component}
\label{sec:halodisp}

\begin{figure*}
\includegraphics{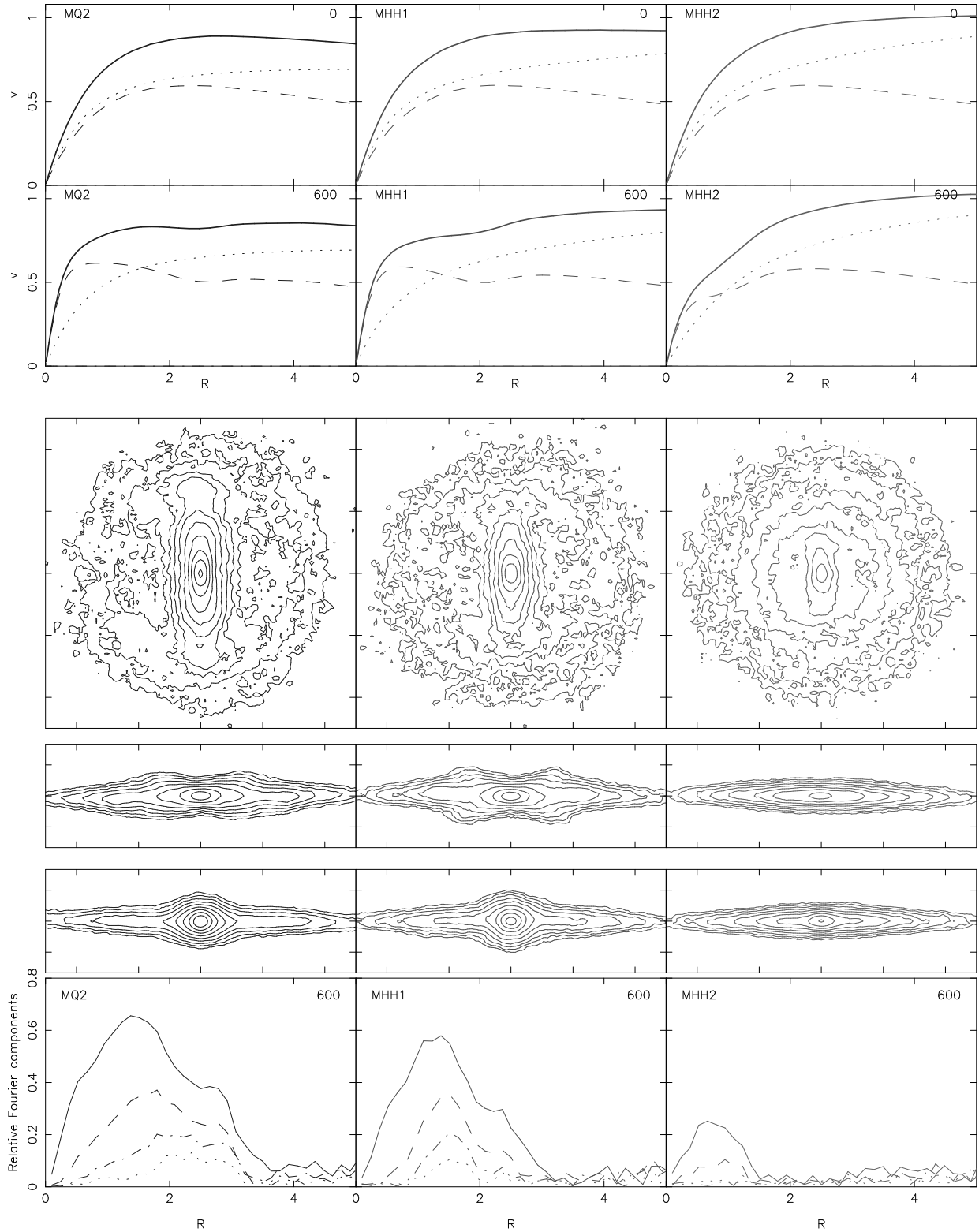}
\vspace{21.2cm}
\caption{Basic information on three simulations with different halo
components, at time $t$ = 600. The layout is as for figure
\ref{fig:basicMh}.   
}
\label{fig:basicHH}
\end{figure*}

\begin{figure}
\includegraphics{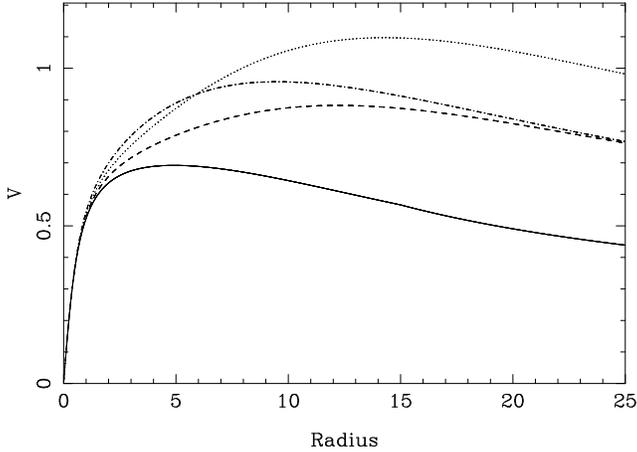} 
\vspace{6.5cm}
\caption{Rotation curve of the initial halo for model
  MQ2 (solid line), MHH1 (dashed line), MHH2 
  (dot-dashed line) and MHH3 (dotted line).
}
\label{fig:h2_f1}
\end{figure}

One of the conclusions of section \ref{sec:halo_analytic} is that, for a
given bar strength, the amount of angular momentum that can be
absorbed by a given halo resonance is larger if the distribution function is
colder. Strictly speaking, this was shown only for haloes whose
distribution function depends on the energy only. It should, however,
be possible to extend this conclusion to a much wider class of
reasonable distribution functions. In this case, and provided the halo
is the main receptacle for the bar angular momentum, colder
haloes should draw more angular momentum from the bars they harbour, than hot
haloes. Since the bar is limited to the region within corotation, it is
a negative angular momentum ``perturbation'', and therefore will grow
stronger in cases where more angular momentum will be be taken from
it, as I already discussed in section \ref{sec:analytic}. I thus come to the
conclusion that colder haloes should harbour stronger bars with 
faster decreasing pattern speed. I will now test these analytical
predictions with numerical simulations.  

The simplest way of increasing or decreasing the velocity dispersion
in the halo component is to stay within the framework described in section
\ref{sec:simulations}, i.e. that of a spherical and isotropic halo, and
increase or decrease the mass of the halo in  
its outer parts. Indeed in that case the halo radial velocity
dispersion can be calculated from the collisionless
Boltzmann equation as  
 
\begin{equation}
\langle u_r^2\rangle ={1\over \rho_{h}(r)}\int_r^{\infty}\rho_{h}\frac
{GM(r')}{r^{'2}}dr',
\label{eq:radialDM}
\end{equation}

\noindent
where $M(r)$ is the cumulative mass distribution. It is easy to see
from the above equation that, for the same or similar $\rho_{h}$, mass
distributions which extend to large radii 
will give larger values of $<u_r^2>$, i.e. hotter haloes.

Let me now compare the results of four simulations, MQ2, MHH1, MHH2 and
MHH3. As given in Table \ref{tab:initcond}, all four have identical
disc mass distributions and $Q$ values. 
The halo rotation curves of three of these simulations at the initial times
are given in the upper panels of Fig.~\ref{fig:basicHH} for the inner
region ($r < 5$ disc scale lengths) and for all four in the left panel of
Fig.~\ref{fig:h2_f1} for distances up to 25 disc scale lengths. The
upper panels of Fig.~\ref{fig:basicHH} also give the disc
and the total rotation curves. Within the inner 10
disc scale lengths the total rotation curve is flat, or, for MHH2,
slightly rising, in all cases compatible with
observed rotation curves. 

Fig.~\ref{fig:basicHH} compares the basic morphological properties of the bar
for three of the simulations, namely MQ2, MHH1 and MHH2. The most
striking difference is between the 
lengths and strengths of the three bars. The former is clearly seen from
the third row of panels, which shows the isophotes of the disc component. In
the leftmost panel the bar length is of the order of 3 to
3.6\footnote{for a summary of the various methods used in measuring
the bar length and the uncertainties involved see section 8 of AM02.} {\it initial} disc
scale lengths, while in the rightmost one it is of the order of 1.5. 
I have given the above estimates in terms of the {\it
  initial} disc scale length, so as to permit direct comparison between
the different cases. Indeed the disc scale length increases with time
(Valenzuela \& Klypin 2002) and the rate of increase could well be
different for the simulations under comparison. Edge-on views 
show that simulations MQ2 and MHH1 have a peanut or `X'-shaped profile 
when viewed edge-on with the bar seen side-on (i.e. with the line of
sight along the bar minor axis). On the other hand simulation MHH2 does
not have any such shape, showing, at the best, a mild boxiness in the
inner parts. 
The lower panels show that the amplitude of all $m$ Fourier components
decreases with increasing 
halo velocity dispersion. Also the location of the maximum moves to
smaller radii as the halo velocity dispersion increases, i.e. as the
bar length decreases.
   
\begin{figure} 
\includegraphics{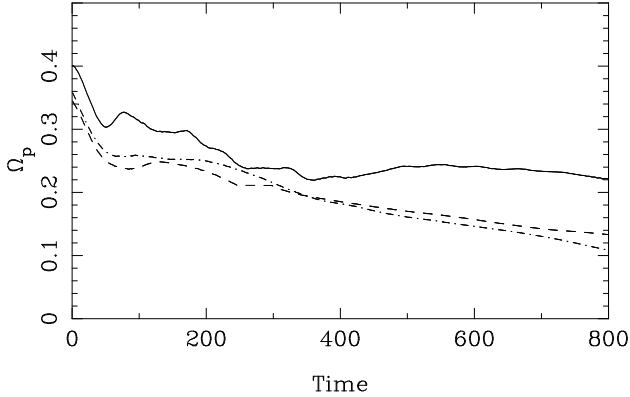} 
\vspace{6.cm}
\caption{Bar pattern speed as a function of time, for simulations
MQ2 (dot-dashed line), MHH1 (dashed line) and MHH2 (solid line). 
}
\label{fig:omega_HH}
\end{figure}

Fig.~\ref{fig:omega_HH} compares the three pattern speeds. It is clear
that, although the pattern speed is a fast decreasing function of time
for simulation MQ2, it hardly decreases for simulation MHH2.
This is due to the fact that the halo in
MQ2 is cold and thus its resonances can absorb a fair amount of angular
momentum. On the contrary in MHH2 the halo is hot and thus
its resonances can not act as an angular momentum sink. This again
shows the importance of angular momentum exchange for the bar slowdown
rate. It is also interesting to note that MHH2 is an example of a
simulation with a strong halo and a hardly decreasing pattern
speed. Thus it argues against a link between relative halo content
and bar slowdown, and, more generally, against using the latter to
set constraints on the former. 

The angular momentum exchange also follows the sequence predicted by
the analytic theory. Thus the angular momentum absorbed by the
halo of simulation MQ2 is 1.6 times that for simulation MHH1, which in
turn is 7 times that of MHH2. Thus this sequence of models behaves as
expected by the analytical results and by the bar formation picture argued
here.

\section{Correlations}
\label{sec:correlations}

Eq. (\ref{eq:DLgeneral}) predicts that the strongest the bar, the
more the disc angular momentum will change.
However, the angular momentum change does not
only depend on the strength of the bar, but also on the distribution
function at the resonant region. Similarly eq. (\ref{eq:DLgeneralh})
shows that the stronger the bar, the more angular momentum the halo 
will absorb. Again, however, the relation depends on the 
distribution function, this time of the halo component.

The relation between the angular momenta of the different components 
and the pattern speed is not as straightforward. The bar angular momentum 
is linked to the pattern speed via eq.~(\ref{eq:Lomp}), 
which includes the moment of inertia of the bar, which in turn depends on the
bar strength and does  
not stay constant with time. Thus the relation of the angular momentum 
change to the pattern speed is more complicated than its relation to the
bar strength. 
 
Checking these relations with the bar angular momentum in the simulations 
is not straightforward, mainly because it is not straightforward to 
define the bar. One can of course attempt to define the outline of the
bar, but several particles move in and out of it and thus are not easy
to classify as belonging to the bar or not. The alternative approach is 
to define the bar as the total of the particles that are trapped around 
the $x_1$, or $x_1$-related orbits. Since these orbits, or their
projections, close after one rotation and two radial oscillations,
this is equivalent to defining the bar as the sum of the particles
that are trapped around the $l$ = -1 $m$ = 2  
resonance. This also, however, is not satisfactory, since bars contain
particles that are trapped at other resonances, as well as chaotic orbits.
I will thus avoid discussing the bar angular momentum and concentrate 
on the halo angular momentum. Since in the simulations presented here
the halo is initially non-rotating, its total angular momentum at any
time will be equal to the angular momentum it has absorbed up to then.
Similarly, since the disc starts as axisymmetric, the bar strength at any
time represents also the global increase of the strength from the
start. 

I will start by following the change of the bar 
strength and pattern speed and the halo angular within a given run, and 
then compare ensembles of runs.

\subsection{Individual simulations}    
\label{subsec:individual} 
 
\begin{figure*}
\includegraphics{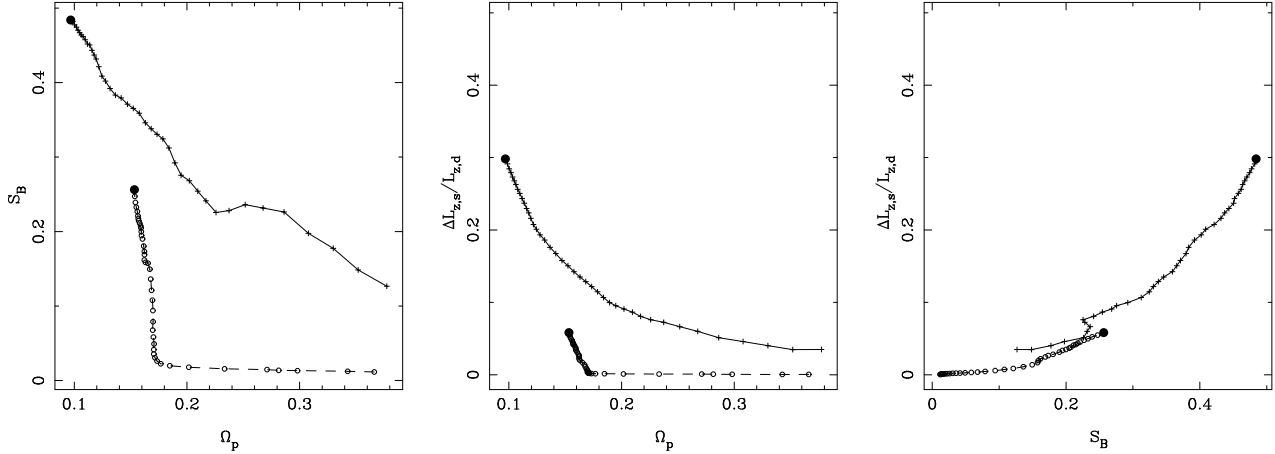} 
\vspace{6.5cm}
\caption{Relations between the bar strength and the pattern speed
(left panel), the halo angular momentum and the pattern speed (middle
panel) and the halo angular momentum and the bar strength (right
panel). The halo angular momentum is normalised to the initial disc
angular momentum ($L_{z,d}$). The times 
plotted start with time 60, to ensure that the bar has grown somewhat,
so that the pattern speed can be measured, and go to the end of 
the simulations, at time 900. The last time is marked with a filled circle. 
The evolution for the simulation with the initially cold disc
($Q_{init}$ = 0.1) is given
with a full line and crosses. That of the initially hot disc  
($Q_{init}$ = 2.2) by a dashed line and open circles.}
\label{fig:evol1}
\end{figure*}

The analytical work in section~\ref{sec:analytic} predicts that, during
the evolution, 
the bar will become stronger and its pattern speed will decrease, 
while the halo angular momentum will increase and that of the disc
decrease. These effects have been qualitatively assessed in many simulations so
far (e.g. Debattista \& Sellwood 1998, 2000; Athanassoula 1996,
2002a,b, 2003). Here I will focus on the relations between these
quantities during the evolution. Fig.~\ref{fig:evol1} 
shows this in a quantitative way for two simulations with 
identical initial conditions, except for the parameter $Q_{init}$, which is
constant with radius, but has a different value in the two
cases. Both simulations have initially $M_d$ = 1, $R_d$ = 1, 
$z_0$ = 0.2, $M_h$ = 5, $\gamma$ = 0.5 and $r_c$ = 10 . Simulation MQ1
has $Q_{init}$  
= 0.1, i.e starts off very cold, and simulation MQ2 has $Q_{init}$ = 2.2,
i.e. starts off very hot. They are extreme cases that show best the 
effect of $Q_{init}$, and have already been discussed in
section~\ref{sec:discdisp}. I have run a set of 8 simulations with
intermediate  
values of $Q_{init}$ and have found intermediate results. I will thus only 
describe the two extremes. Note that their evolution is quantitatively
very different. 

For the initially cold simulation both the pattern speed and the bar 
strength change is a similar way, so that their relation is simple 
and could, to a first approximation, be described by a single straight 
line in the $(S_B,\Omega_p)$ plane. This is not the case for the
initially hot simulation, which has  
two distinct evolution phases. For this simulation
the bar strength hardly increases during the first part of the evolution, 
while, during this time, the pattern speed decreases
drastically\footnote{Note that during this stage the pattern speed is 
poorly defined.}. Then, very abruptly, the situation changes 
and the pattern speed stops decreasing, while the 
strength of the bar starts increasing. The relation between the two
quantities and the halo angular momentum is clear in the next two frames. 
Again for the initially cold disc the evolution is more gradual, 
while for the initially hotter one it is in two episodes. 

\subsection{Some global trends}
\label{subsec:global}

\begin{table}
\caption[]{Correlation coefficients}
\begin{flushleft}
\label{tab:correl}
\begin{tabular}{rrrrrrr}
\hline
 & \multicolumn{2}{|c|}{$(S_B,\Omega_p)$} &
 \multicolumn{2}{|c|}{$(\frac{\Delta L_{z,s}}{L_{z,d}},\Omega_p)$} &
 \multicolumn{2}{|c|}{$(\frac{\Delta L_{z,s}}{L_{z,d}},S_B)$} \\
\hline
 & $N_{s}$ & $r_{cor}$ & $N_{s}$ & $r_{cor}$ & $N_{s}$ & $r_{cor}$ \\
\hline
All simulations             & 125 & -0.83 & 116 & -0.76 & 116 & 0.95 \\
.8~$\le~Q_{init}~\le $~1.2 &  54 & -0.88 &  51 & -0.89 &  53 & 0.97 \\
$Q_{init} <$ 0.3            &  13 & -0.88 &  13 & -0.83 &  13 & 0.99 \\
with bulge                  &  13 & -0.96 &  13 & -0.90 &  13 & 0.95 \\
$\gamma <$ 2                &  49 & -0.74 &  51 & -0.89 &  48 & 0.84 \\
0.1 $\le \gamma \le$ 0.5    &  40 & -0.79 &  41 & -0.94 &  38 & 0.88 \\
$\gamma =$ 0.01             &   3 & -0.87 &   - &   -   &   - &  -   \\
$\gamma \ge$ 5              &  46 & -0.74 &  39 & -0.55 &  39 & 0.91 \\
$\gamma \ge$ 5, cold        &  40 & -0.63 &  33 & -0.42 &  33 & 0.88 \\
$\gamma \ge$ 5, hot         &   6 & -0.90 &   6 & -0.74 &   6 & 0.94 \\
\noalign{\smallskip}
\hline
\end{tabular}
\end{flushleft}
\end{table}

\begin{figure*}
\includegraphics{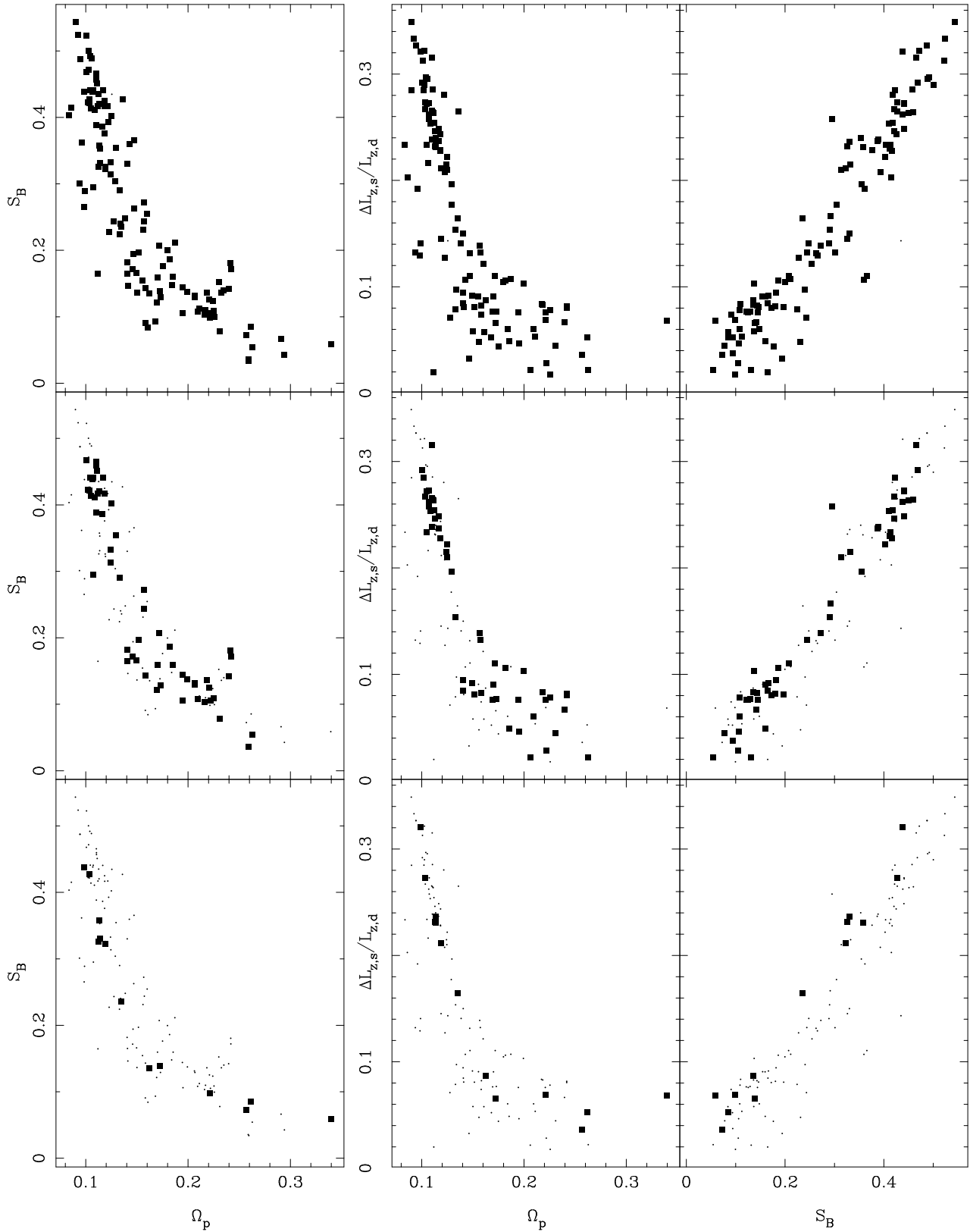} 
\vspace{21.2cm}
\caption{Relations between the bar strength and the pattern speed
(left panels), the spheroid angular momentum and the pattern speed (middle
panels) and the spheroid angular momentum and the bar strength (right
panels), at times $t = $ 800. The spheroid angular momentum is
normalised by the initial disc angular momentum ($L_{z,d}$). The
simulations under consideration in each panel are
marked with a filled solid square and the rest by a dot. The upper row
includes most simulations, the middle one those with 
0.8 $\le Q_{init} \le$ 1.2 and the lower one those with $Q_{init} <$ 0.3.
}
\label{fig:correl1}
\end{figure*}

\begin{figure*}
\includegraphics{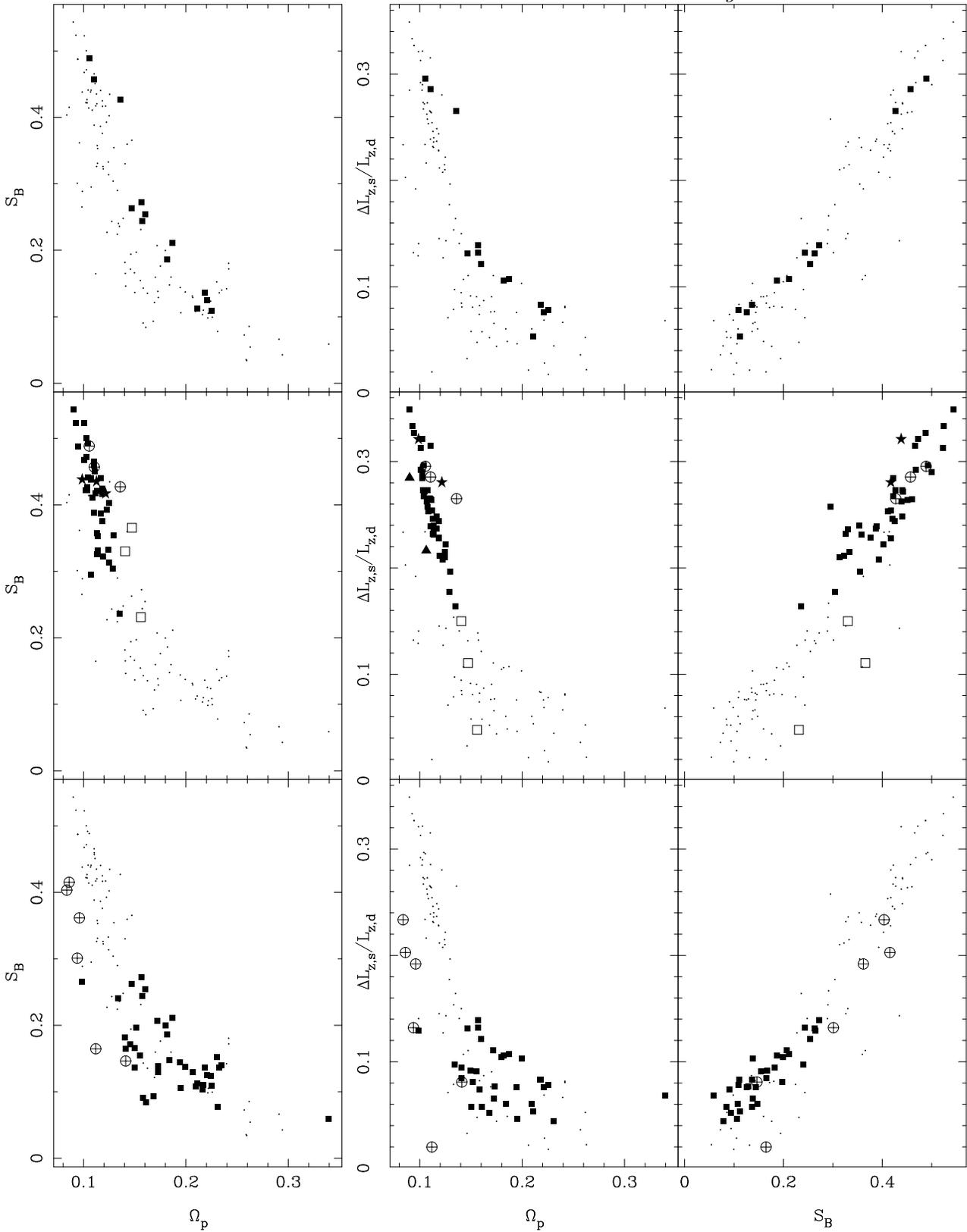} 
\vspace{21.2cm}
\caption{Same as the previous figure but for different cases. The upper
  row includes simulations with bulges, the middle one simulations
  with a halo with $M_{h2}$ = 0, $M_{h1}$ = 5 and $\gamma <$ 2 and the lower one
  simulations with a halo with $M_{h2}$ = 0, $M_{h1}$ = 5 and $\gamma >$ 2. In the
  middle panel simulations with a bulge are marked with a $\oplus$,
simulations with $\gamma =$ 0.01 with a filled star, simulations with  
$\gamma \ge$ 1 with a filled triangle and simulations with $Q_{init}
\ge$ 2 with an open square. In the lower panel simulations with
$Q_{init} \ge$ 1.4 and $z_0 \ge$ 0.2 are marked with an $\oplus$. 
}
\label{fig:correl2}
\end{figure*}

Let me now turn to a more global view of the simulations. According to
what has been said so far, I expect to find a trend between the
strength and pattern speed of the bar and the angular
momentum of the spheroidal component (i.e. halo plus bulge) of all
models. These quantities are plotted in Figs.~\ref{fig:correl1} and
\ref{fig:correl2}, while the corresponding correlation coefficients are
given in Table~\ref{tab:correl}. These are only given as indicative,
and should not be taken too strictly. Indeed, the initial conditions
of the various simulation were not taken so as to cover uniformly the
available parameter space. Rather, they were made so as to follow
interesting results, and these may occur in restricted areas of
the parameter space. Nevertheless, seen the large number of available
simulations, the correlation coefficients should definitely give
indications of trends and correlations. In each case, I list first the
number of relevant simulations, $N_{s}$, and the the corresponding
correlation coefficient.

The upper row in figures~\ref{fig:correl1} and \ref{fig:correl2}
includes all simulations available by the end of July 2002, except
for some simulations which were run for test purposes,
e.g. with a very low number of particles or a very large
softening and a few for which some data were accidentally lost. 

Eq.~(\ref{eq:DLgeneralh}) leads us to expect a very tight
relationship between the spheroidal angular momentum and the bar
strength. This prediction is indeed borne out by the upper panel of
Fig.~\ref{fig:correl1}. Table~\ref{tab:correl} gives for this relation
a correlation coefficient of 0.95. This is a high value, particularly
if one takes 
into account the fact that the plotted quantities (and particularly
the bar strength) have measurement errors and that
eq.~(\ref{eq:DLgeneralh}) includes e.g. the halo distribution function,
which could differ from one case to another. 

Section~\ref{sec:analytic} predicts also trends for the bar pattern speed.
The upper row in Fig.~\ref{fig:correl1} confirms this
expectation. There is an anti-correlation between the bar
strength and pattern speed, albeit less tight than with the spheroid
angular momentum, while the relation
between the spheroid angular momentum and the pattern speed seems to have
two distinct parts. These will be discussed further below.

The remaining panels of Figs.~\ref{fig:correl1} and \ref{fig:correl2}
use only subsamples of simulations. Indeed for such subsamples I
expect to have more homogeneity of e.g. the
distribution function which influences the relationships between the three
quantities.

The middle row of panels concentrates on simulations which have $Q_{init}$
around 1. I have here retained only simulations with 
0.8~$\le~Q_{init}~\le $~1.2. As can be seen from the figure and the
corresponding coefficients in Table~\ref{tab:correl}, the correlations
and trends have, as expected, tightened considerably. 
The lower row of panels focuses on simulations with $Q_{init} <$ 0.3,
i.e. initially very cold. Now the
correlation between the spheroid angular momentum and the bar strength has
reached a correlation coefficient of 0.99. 

The upper row of panels of Fig.~\ref{fig:correl2} includes only simulations
with bulges. They also form correlations, with the $(S_B, \Omega_p)$
and $(\Delta L_{z,s}/L_{z,d},S_B)$ ones being rather strong
(correlation coefficient of -0.96 and 0.95, respectively). A
closer look shows that they do not cover exactly the same area as the
simulations without bulges. This is particularly clear in the $(S_B,
\Omega_p)$ case, where they lie above and to the right of the area
covered by the remaining cases. This displacement is particularly
strong for simulations with large bulges. 

The middle row of panels includes only simulations with live haloes with
$M_{h2}$ = 0, $M_{h1}$ = 5 and a small core radius ($\gamma <$ 2) i.e. 
MH-type simulations. They present tight correlations, particularly those
including the spheroid angular momentum. To examine this further I
plot with different symbols 
simulations with $\gamma <$ 0.02 (filled stars), simulations with
$\gamma >$ 0.9 (filled triangles) and simulations with bulges (crosses
within an open circle). All the remaining simulations have
$\gamma =$ 0.5, except for one which has $\gamma =$ 0.1. They
show a very tight anti-correlation in the $(\Delta L_{z,s}/L_{z,d}, \Omega_p)$
plane (correlation coefficient of -0.94). This is predicted in the
discussion following eq. (\ref{eq:Lomp}), where a correlation
is predicted between the spheroid angular momentum and the bar
pattern speed, if the main receptacle of angular momentum is the
halo. This should indeed be the case for MH-type simulations. Thus the
tight anti-correlation found is yet another argument in favour of the bar
evolution picture proposed here. 

Finally the lower panels correspond to cases with $M_{h2}$ = 0, $M_{h1}$ = 5 and
a large core radius ($\gamma \ge$ 5), i.e. to MD-type cases. Their
behaviour is totally different from that of the MH-type cases. The
only correlation worth mentioning is the one between the spheroid angular
momentum and the bar strength. The striking difference between the cases with
a small core radius ($\gamma <$ 2, middle panels) and those with a large
core radius ($\gamma \ge$ 5, lower panels) could be predicted from the
framework presented here. Indeed in the cases with a small core radius
the halo is the main recipient of angular momentum, as discussed
above, and there should be correlations between the angular momentum
it has acquired and the bar properties. On the other hand for cases
with a large core radius, then the outer disc may also absorb a
non-negligible part of the angular momentum, and thus a correlation with
the halo angular momentum only would be less tight. This is better
understood if we consider 
separately in the lower panels the cases with hot discs,
i.e. $Q_{init} \ge$ 1.4 and $z_0 \ge$ 0.2 (crosses
within an open circle), and the colder cases (filled squares). It is
interesting to note that they form 
separate, clear sequences in all three planes, as could be expected
from the scenario I propose
here. Indeed if the outer disc is hot, then the spheroid should
be the main angular momentum sink and thus trends or correlations with
the spheroid 
angular momentum should be expected. On the other hand if the outer
disc is cold, then both it and the spheroid should be sinks of angular
momentum, and a trend or correlation with the angular momentum
absorbed by only one of the two will not necessarily exist, or will be
more loose. All this is well borne out by the relations in the lower
panels.

\section{Summary and discussion}
\label{sec:sumdisc}

In this paper I discussed the role of angular momentum exchange in
determining the length, the strength and the slowdown
rate of bars. In order for the bar to grow, it has to shed
angular momentum, and, for that, there must be resonant material to
receive it. Such material can be found either in the outer disc, or in
the halo. It is this exchange that drives the bar evolution and
determines its strength and length as well as its slowdown. 

The galaxy strives to transfer angular momentum outwards (LBK), while
keeping an equilibrium between emitters and absorbers. Emitting can
be hampered by increasing the disc velocity dispersion within CR 
(section \ref{sec:discdisp}). There will then be only little
angular momentum exchange -- even if the halo has the possibility of
absorbing considerable amounts -- and this will limit the pattern speed
decrease and the bar strength. Absorbing by the outer disc can be
hampered if that part of the disc has low density, or is hot (section
\ref{sec:discdisp}). Absorbing by the spheroid is hampered if that
component is hot, or rigid (section \ref{sec:halodisp}), or has
relatively little mass in the resonant regions (section
\ref{sec:halodens}). More angular momentum exchange leads to stronger
bars, whose pattern speed decreases faster. To make
the role of the angular momentum exchange yet clearer I presented
trends and correlations between the exchanged angular momentum and the
bar strength and pattern speed.

As discussed in section \ref{sec:orbits}, in order to shed angular
momentum the bar can slow down, or become longer, or thinner. The
$N$-body simulations have shown that it chooses to follow all three
alternatives, but to a degree that varies from one simulation to
another and also sometimes during different phases of the
evolution. More work is needed to explain this behaviour. 

If the galaxy is isolated, there
should be as much angular momentum emitted as there is absorbed,
i.e. there should be an equilibrium between emitters and absorbers. 
This equilibrium influences the evolution of the pattern speed.
Indeed, if the pattern speed is lowered, corotation will
move outwards and there will be more emitters and less disc
absorbers. Thus in galaxies in which the halo can not absorb much
angular momentum,
either because of its low density or because of its high velocity
dispersion, corotation should be at a relatively small radius to make space
for sufficient absorbers in the disc component. On the other
hand, in galaxies with a strong and responding halo component,
corotation can be at larger radii. This will maximize the number of
emitters, while the absorption will be assured mainly by the
halo. These predictions are indeed borne out by the simulations
presented here. Indeed MH-type
simulations have longer bars than MD-types. Also the
pattern speed of MH-type bars decreases faster than that of MD-types.

The considerations in this paper argue that it is very hazardous
to set limits to the relative halo mass from the bar slowdown rate. I have
shown here that this slowdown depends not only on the relative halo
mass, but also on the velocity dispersion of both the disc and the
halo (bulge) component. Indeed, in sections~\ref{sec:discdisp},
\ref{sec:density},and  
\ref{sec:halodisp} I gave examples showing the crucial importance 
of these parameters. Models MH3 and MHH2 have a very
strong halo. Yet their pattern speed hardly decreases.
A similar comment can be made for model
MQ8. Furthermore, the reason for this is not the same in all cases. In
model MHH2 the halo is too hot to be 
able absorb, while in model MQ8 the disc is very hot, thus
hampering both emitters and disc absorbers. In all these cases the
pattern speed decrease is of the order of, or less than, 0.005 in a
$\Delta t$ of 100 computer units. Applying the calibration proposed in
AM02, I find that this corresponds to a decrease of the pattern speed
of the order of, or less than, 0.35 km/sec/kpc in 1.4 Gyrs. This
decrease is considerably less than the observational errors when
measuring the pattern speed (see e.g. Gerssen 2002 for a compilation).
It thus can not be excluded that disc
galaxies have a high halo-to-disc mass ratio and at the same time a hardly
decreasing pattern speed. Let me continue this argument further and
assume that all bars were in place at about $z$ = 0.5, i.e. 2 to 3 Gyrs
ago. Then, for galaxies ressembling simulations MHH2 or MQ8, the bar
pattern speed would have decreased by less than 1
km/sec/kpc since that time, again considerably less than
observational errors, while their halo would be relatively very massive. One
should thus not use the pattern speed slowdown as a way of setting a
limit to the halo-to-disc mass ratio. Several other ways
have been put forward for the case of barred galaxies (Athanassoula
2002b) and they will be discussed in some detail in a future paper.  

Although the picture is now sufficiently clear for $N$-body bars, it is
not straightforward to apply it to real galaxies. Of course, the trends I have
found here should carry over. In other words, both the outer disc
and the halo should be able to absorb more angular momentum if the
resonant regions are more densely populated and/or if the resonant
material is colder. Our knowledge, however, of the halo properties is
rather limited, and does not allow us to go much further. The halo
material could be elementary particles, or baryonic material of
sub-stellar masses, or small black holes. In these cases the mass
of each `halo particle' ranges from smaller to {\it very} much smaller than
the mass of the individual particle used in the present
simulations. It should be sufficiently small to ensure
that galactic haloes have low graininess and are thus far from the
noise-dominated regime. Hence there should be considerable halo
material at or near resonance, able to absorb angular momentum. The
angular momentum exchange should thus be
particularly strong, much more so than in the present simulations, in
the case of a halo made of elementary particles 
having a relatively cold distribution function. On the other hand, the
sub-clumps which could well exist in the halo would lower the capacity
of the halo to absorb
angular momentum to a level more comparable to what is found in the
present simulations.

The mass and the density distribution of the halo material in the
regions of interest is not well known, as witnessed by the fact that 
the debate between the maximum disc and the sub-maximum disc
proponents is still 
going strong (see e.g. Bosma 1999 and 2002, and references therein). Even
less is known about the axial ratio of the halo, about whether it has
figure rotation, about its extent, and 
about how much, if at all, the material in it rotates. Finally, it is impossible
to say anything about even basic properties of the halo distribution function.

Seen our very restricted knowledge about the halo properties, it might
be useful to face the question from the opposite direction,
i.e. to see what, if anything, these simulations can tell us about the
halo. Unfortunately no strong conclusions can be obtained, although
there are some suggestions and indications. Namely galaxies with
strong bars should have offered a sufficient sink of angular momentum
so that the bar could grow to its present strength. On the other hand
weak bars must be in surroundings where the angular momentum exchange
was limited. This could be either by a very extended, or otherwise hot,
halo, or then a halo whose resonance regions are of low
density. 

The results presented in this paper explain well the observations
discussed in the introduction. Thus bars can come in a broad range
of strengths because the amount of angular momentum exchanged in their
respective galaxies varies considerably from one case to another. In
galaxies with weak, small 
bars -- as e.g. our own Galaxy -- little angular momentum should have been
exchanged. The contrary should be true for galaxies with very strong
bars. In this context it is worth mentioning that Gadotti and de
Souza (2003a, 2003b) report on two galaxies with particularly strong
bars. Most of the light in these two galaxies is in the bulge/bar
component and very little in the disc. Making the reasonable
assumption that the mass-to-light ratio of the two have similar values,
I come to the conclusion that the bar has grown sufficiently in those
galaxies to take over most of the material originally in the disc, and
leave a very weak disc component. These galaxies
would then be extreme cases of MH-type galaxies.

The correlation between bar lengths and bulge sizes, reported by
Athanassoula \& Martinet (1980) and by Martin (1995), can also be
understood in this framework. Indeed the bulge is part of the
spheroidal component and thus can
help the bar grow by taking from it angular momentum. The more massive
the bulge, the more angular momentum it can absorb and the longer the
bar will grow. Finally, this
framework explains why bars are stronger in early type galaxies than
in later types. Indeed these galaxies are known to have stronger
bulges, which will act as angular momentum sinks and thus simulate bar
growth. Whether one can
extend this argument further and deduce that early types should also
have haloes which can act as a better sink of angular momentum than
the haloes of late type galaxies, is not
clear at this point. Detailed modelling of a few barred galaxies would
be necessary to estimate how much angular momentum their bulges can
take and thus to see to what extent the haloes of early type galaxies
need to absorb more angular momentum than their counterparts in late
types.

Rings are often observed in disc galaxies (e.g. Buta 1995, 1999) and their
location is known to be linked to resonances (Athanassoula, Bosma,
Cr\'ez\'e et al. 1982; Buta 1995, 1999). If the pattern speed decreases
considerably with time the position of the rings will migrate
outwards, while if the pattern speed hardly changes the ring position
will also not change much. It would thus be interesting to pursue a
detailed modelling, including also gas, star formation and stellar
evolution, in order to see whether there could be observable
signatures from ring migration on the population of the ring. It would
then be possible to set constraints on the evolution of a disc galaxy
from spectral observations of its ring(s).

Several simulations in the last few years have formed bars of very different
length, strength and slowdown rate. These differences can now be
easily understood in the framework presented here. Debattista \&
Sellwood (1998, 2000) reported results from relatively low resolution
simulations (i.e. with a softening length equal to one fifth of the disc
scale length and to twice the disc scale height),  
which start off very cold ($Q_{init}$ = 0.05), while their
haloes had a relatively short extent (12.6 disc scale lengths). My
results here show that all their MH-type simulations
should result in strong bars with sharply decreasing pattern
speed, and indeed this is what their simulations gave. AM02, A02 and
Athanassoula (2002a) presented simulations with a considerably smaller
softening (in most cases 0.0625 of the disc scale length, or,
equivalently, less than a third of the disc scale height; in some
cases 0.03125), a wide range of $Q_{init}$ 
and a somewhat more extended halo (15 disc scale lengths). These
results are discussed in detail in this paper, and fit 
well the bar evolution picture presented here. Finally Valenzuela \&
Klypin (2002) have three simulations with  $Q_{init}$ higher than 1, and a
very extended halo component (70 to 85 disc scale lengths). Their
softening is variable and, at 
least in the innermost parts, very small. It is not clear at this
stage whether the number of particles in the disc (200\,000) and the inner halo
is sufficiently large to prevent noise from influencing the
evolution with this small a softening. The general evolution,
however, is in good agreement with 
the picture presented here. Indeed, they report very little
angular momentum exchange and bar slowdown rate, and their bars are
somewhat shorter than those e.g. of AM02. This could be expected
since their halo particles are too hot to absorb considerable angular
momentum. Thus their results are to be compared with e.g. model MHH2 of the
present paper.
 
The picture presented in this paper is rather complicated, since it includes
various sinks (the material at resonance in the outer parts of the
disc, as well as the halo resonant material) and is influenced by
several properties of 
the galaxy. Even so, this picture is not complete and includes several major
simplifications. One concerns the distribution function of the
halo. Here I assumed 
a particularly simple case. However, since the form of the
distribution function will influence the amount of angular momentum
exchanged and therefore the strength and slowdown of the bar, it is
interesting to examine other distribution functions to find what range of angular
momentum exchanges can be covered. A second important point to
consider is the existence of a gaseous component. This would be one
more partner in the angular 
momentum exchange, complicating further the problem. Indeed, even if
the gas mass is only a very small fraction of the total, it is very
cold, considerably colder than both the disc and the halo. Thus its
contribution to the angular momentum exchange process could well not
be negligible. 

A third effect is the shape of the 
halo. Although the flattening of the halo could influence the
angular momentum exchange, in particular via the halo distribution
function, I believe that the strongest effect would come from the
possible non-axisymmetry of the halo. By this I do not mean the
possible elongation of the halo in the inner region in response to the
bar, which is anyway included in the simulations discussed here, since
they are fully 
self-consistent. I mean the effect that a non-axisymmetry, present in
the halo at large scale in the initial conditions, could have on the
evolution in general. In such a case the halo and bar components could interact
in a way reminiscent of mode-mode interaction often discussed in
plasma physics, the one driving the other. Since cosmological
simulations show the existence of strongly 
triaxial haloes, I will be addressing this very complicated problem in
a future paper. A fourth related issue is the effect of companions,
since galaxies are often not isolated, and thus a companion could be
one more partner in the angular momentum exchange process. A first
attempt at this problem is made by Berentzen et al. (2003).

A possible objection against the simulations presented here is that
bars probably do not form in as quiet a way as that described
here. Indeed here I assumed, as bar formation studies always do,
that the disc is initially axisymmetric, with properties near to
those of present day disc galaxies. In other words I assumed
implicitly that the disc formed first and the bar later. This would be
a reasonable hypothesis if bars are rare at high redshift, as argued
by Abraham et al. (1999). If this is not the case and bars start
growing during the disc 
formation stage, then the detailed properties of the bar may be
considerably different from those of the bars presented
here. Nevertheless, this  
will not change any of the physics described here. Again it
will be the angular momentum exchange that will influence the bar
growth and slowdown, although the differences in the properties
of the disc and halo from those assumed here may well change the
{\it amount} of angular momentum exchanged, and thus the detailed bar
properties. Since, however, I have studied here both qualitatively and
quantitatively the effect of the disc and halo
properties on the angular momentum exchanged, it should be possible,
when the subject of disc formation history is more evolved, to apply the
results found here to the proper disc formation scenario.   

\parindent=0pt
\def\rr{\par\noindent\parshape=2 0cm 8cm 1cm 7cm}
\vskip 0.7cm plus .5cm minus .5cm

{\Large \bf Acknowledgments.} I thank M. Tagger, A. Bosma,
W. Dehnen, C. Heller, I. Shlosman, F. Masset, J. Sellwood, O. Valenzuela,
A. Klypin, P. Teuben and A. Misiriotis for many interesting discussions. 
I thank J.~C. Lambert for his help with the GRAPE
software and the administration of the simulations and W. Dehnen for
making available to me his tree code and related programs. I
also thank the INSU/CNRS, the University of Aix-Marseille I, the Region PACA
and the IGRAP for funds to develop
the GRAPE and Beowulf computing facilities used for the simulations
discussed in this paper and for their analysis. 
\vskip 0.5cm

{\Large \bf References.}
\rr{Abraham, R. G., Merrifield, M. R., Ellis, R. S., Tanvir, N. R. \&
Brinchmann, J. 1999, \MN, 308, 569}
\rr{Athanassoula, E. 1992, \MN, 259, 345}
\rr{Athanassoula, E. 1996, in ``Barred Galaxies'', eds. R. Buta,
D. Crocker and  B. Elmegreen, PASP conference series, 91, 309}
\rr{Athanassoula, E. 2002a, \ApJ, 569, L83 (A02)}
\rr{Athanassoula, E. 2002b, in ``Galactic Discs : Kinematics, Dynamics and
Perturbations'', eds. E. Athanassoula, A. Bosma and R. Mujica, PASP
conference series, 275, 141}
\rr{Athanassoula, E. 2003, in ``Galaxy Evolution: Theory
and Observations'', eds. V. Avila-Reese, C. Firmani, C. Frenk and C.
Allen, RevMexAA, in press}
\rr{Athanassoula, E., Bienaym\'{e}, O., Martinet, L. \& Pfenniger,
D. 1983, \AAA, 127, 349}
\rr{Athanassoula E., Bosma, A., Cr\'ez\'e, M. \& Schwarz, M. P. 1982, \AAA, 107, 101}
\rr{Athanassoula, E. \& Martinet, L. 1980, \AAA, 87, L10}
\rr{Athanassoula, E. \& Misiriotis, A. 2002, \MN, 330, 35 (AM02)}
\rr{Athanassoula E. \& Sellwood J. A. 1986, \MN, 221, 213}
\rr{Berentzen, I., Athanassoula, E., Heller, C. H. \& Fricke, K. J. 2003,
in preparation}
\rr{Binney, J. \& Tremaine, S. 1987, Galactic Dynamics, Princeton
University press}
\rr{Bosma, A. 1999, Celestial Mechanics \& Dynamical Astronomy, 72, 69}
\rr{Bosma, A. 2002, in ``Galactic Discs : Kinematics, Dynamics and
Perturbations'', eds. E. Athanassoula, A. Bosma and R. Mujica, PASP
conference series, 275, 23}
\rr{Buta, R. 1995, \ApJS, 96, 39}
\rr{Buta, R. 1999, Ap\&SS, 269/270, 79}
\rr{Contopoulos G. 1980, \AAA, 81, 198 }
\rr{Contopoulos G., Grosb{\o}l P. 1989, A\&AR 1, 261}
\rr{Contopoulos G., Papayannopoulos, T. 1980, \AAA, 92, 33}
\rr{Debattista, V. P., \& Sellwood, J. A. 1998, \ApJ, 493, L5}
\rr{Debattista, V. P., \& Sellwood, J. A. 2000, \ApJ, 543, 704}
\rr{Dehnen, W. 2000, \ApJ, 536, L39}
\rr{Dehnen, W. 2002, J. Comp. Phys., 179, 27}
\rr{Elmegreen, B.G., Elmegreen, D. M., 1985, \ApJ, 288, 438}
\rr{Gadotti, D. A. \& de Souza, R. E. 2003a, ApJL, 583, L75}
\rr{Gadotti, D. A. \& de Souza, R. E. 2003b, in ``The evolution of
galaxies III From simple 
approaches to self-consistent models'', eds. G. Hensler, G. Stasinska,
S. Harfst, P. Kroupa, Chr. Theis, Kluwer academic
publishers, in press}
\rr{Gerssen, T. 2002, in ``Galactic Discs : Kinematics, Dynamics and
Perturbations'', eds. E. Athanassoula, A. Bosma and R. Mujica, PASP
conference series, 275, 197}
\rr{Hernquist, L. 1993, \ApJS, 86, 389}
\rr{Hernquist, L. \& Weinberg, M. D. 1992, \ApJ, 400, 80}
\rr{Kalnajs, A. J. 1971, \ApJ, 166, 275}
\rr{Kato, S. 1971, \PASJ, 23, 467}
\rr{Kawai, A., Fukushige, T., Makino, J., \& Taiji, M. 2000,
  \PASJ, 52, 659}
\rr{Kormendy, J. 1979, \ApJ, 227, 714}
\rr{Little, B. \& Carlberg, R. G. 1991a, \MN, 250, 161}
\rr{Little, B. \& Carlberg, R. G. 1991b, \MN, 251, 227}
\rr{L\"utticke, R., Dettmar, R.-J. \& Pohlen, M. 2000, \AAA, 362, 435}
\rr{Lynden-Bell, D. \& Kalnajs, A. J. 1972, \MN, 157, 1, 1972 (LBK)}
\rr{Mark, J. W.-K. 1976, \ApJ, 206, 418}
\rr{Martin, P. 1995, \AJ, 109, 2428}
\rr{Ohta, K. 1996, in ``Barred Galaxies'', eds. R. Buta,
D. Crocker and  B. Elmegreen, PASP conference series, 91, 37}
\rr{Reynaud D. \& Downes D. 1997 \AAA 319, 737}
\rr{Sandage A. 1961, The Hubble Atlas of Galaxies, Carnegie
Institution of Washington, Washington DC}
\rr{Sandage, A. \& Bedke J. 1988, Atlas of Galaxies, NASA, Washington DC}
\rr{Sanders R. H., Tubbs A. D. 1980 \ApJ, 235, 803}
\rr{Sellwood, J. A. 1980, \AAA, 89, 296}
\rr{Skokos, H., Patsis, P. \& Athanassoula, E. 2002,
\MN, 333, 861, 2002}
\rr{Sygnet, J. F., Tagger, M., Athanassoula, E. \& Pellat, R. 1988,
\MN, 232, 733}
\rr{Tagger, M., Sygnet, J. F., Athanassoula, E. \& Pellat, R. 1987,
\ApJ, 318, L43}
\rr{Tremaine, S. \& Weinberg, M. D. 1984a, \MN, 209, 729}
\rr{Tremaine, S. \& Weinberg, M. D. 1984b, \ApJ, 282, L5}
\rr{Weinberg, M. D. 1985, \MN, 213, 451}
\rr{Valenzuela, O. \& Klypin, A. 2002, MNRAS, submitted}
\rr{Van Albada, T. S. \& Sanders, R. H. 1982, MNRAS, 201, 303}

\label{lastpage}
\end{document}